\documentclass[fleqn,usenatbib]{mnras}
\usepackage{newtxtext,newtxmath}
\usepackage[T1]{fontenc}
\usepackage[utf8,latin1]{inputenc}
\usepackage{graphicx}	
\usepackage{amsmath}	
\usepackage{epsfig}
\usepackage{setspace}
\usepackage{graphicx}
\usepackage{siunitx}
\usepackage{xcolor}
\usepackage{hyperref}
\hypersetup{colorlinks, linkcolor={red},citecolor={blue},urlcolor={blue}}
\usepackage{epstopdf}  
\bibpunct{(}{)}{;}{a}{}{,}
\usepackage[english]{isodate}
\usepackage{aas_macros}
\usepackage{array}   
\newcolumntype{L}{>{$}l<{$}}
\newcolumntype{C}{>{$}c<{$}}

\def\gro{GRO~J1655-40}
\def\be{\begin{equation}} 
\def\ee{\end{equation}}
\def\fekal{Fe $K_{\alpha}$ line profiles}
\def\tbabs{\texttt{TBabs}}
\def\pexriv{\texttt{pexriv}}
\def\diskbb{\texttt{diskbb}} 
\def\relline{\texttt{relline}}

\usepackage{color}

\title[Exploring HOIs with Fe $K_{\alpha}$-lines from relativistic disks]{Exploring higher order images with Fe $K_{\alpha}$-lines from relativistic disks: black hole spin determination and bias}

\author[Falanga et al. (2020)]{
{M. Falanga,$^{1}$\thanks{mfalanga@issibern.ch} P. Bakala,$^{2,3}$ R. La Placa,$^{4,5,6}$ V. De Falco,$^{3}$, A. De Rosa,$^7$ and L. Stella$^{4}$}\\   
$^{1}$ International Space Science Institute (ISSI), Hallerstrasse 6, 3012 Bern, Switzerland\\
$^{2}$ Research Centre for Computational Physics and Data Processing, Silesian University in Opava, Bezru\v{c}ovo n\'am.~13, CZ-746\,01, Opava, Czech Republic\\
$^{3}$ M. R. \v{S}tef\'anik Observatory and Planetarium, Sl\'adkovi\v{c}ova 41, 920 01 Hlohovec, Slovak Republic\\
$^{4}$ INAF--Osservatorio Astronomico di Roma, via Frascati 33, 00078 Monteporzio Catone (Roma), Italy\\   
$^{5}$ Dipartimento di Fisica, Universit\`a di Roma ``Tor Vergata'', Via della Ricerca Scientifica, 00133 Roma, Italy\\   
$^{6}$ Dipartimento di Fisica, Universit\`a La Sapienza, Piazzale Aldo Moro 5, 00185 Roma, Italy\\ 
$^{7}$ INAF - Istituto di Astrofisica e Planetologie Spaziali, Via Fosso del Cavaliere, 00133 Roma, Italy
} 

\date{Received \today} 
\pubyear{2020}

\begin{document}
\label{firstpage}
\pagerange{\pageref{firstpage}--\pageref{lastpage}}
\maketitle

\begin{abstract}
We study the contributions to the relativistic Fe $K_{\alpha}$ line profile from higher order images (HOIs) produced by strongly deflected rays from the disk which 
cross the plunging region, located between the innermost stable circular orbit (ISCO) radius and the event horizon of a Kerr black hole. 
We investigate the characteristics features imprinted by the HOIs in the line profile 
for different black hole spins, disk emissivity laws and inclinations. We find that they extend from the red wing of the profile up to energies slightly lower than those of the blue peak, adding $\sim 0.4 - 1.3$\% to the total line flux. The contribution to the specific flux is often in the $\sim 1$\% to 7\% range, with the highest values attained for low and negative spin ($a\lesssim 0.3$) black holes surrounded by intermediate inclination angle ($i\sim40^{\circ}$) disks. 
We simulate future observations of a black hole X-ray binary system with the Large Area Detector of the planned X-ray astronomy \emph{enhanced X-ray Timing and Polarimetry Mission} (eXTP) and find that the \fekal\ of systems accreting at $\lesssim 1 $\% the Eddington rate are affected by the HOI features for a range of parameters. 
This would provide evidence of the extreme gravitational lensing of HOI rays. 
Our simulations show also that not accounting for HOI contributions to the Fe $K_{\alpha}$ line profile may systematically bias measurements of the black hole spin parameter towards values higher by up to $\sim 0.3$ than the inputted ones.
\end{abstract}


\section{Introduction}  
\label{sec:intro}
Astrophysical Black Holes (BHs) as described by General Relativity (GR) are remarkably simple objects characterized by just their mass, $M$, and angular momentum, $J$ (their electric charge $Q$ is assumed to be negligibly small). BH angular momentum, often expressed in terms of the dimensionless spin parameter $a \equiv Jc/GM^2$ with $|a|\leq 1$, induces frame-dragging, the rotation of the spacetime around BHs, which affects considerably the motion of matter and light in the close vicinity of the event horizon and enables extraction of BH rotational energy \citep{Misner1973}. BH spin is of great importance across the whole mass scale, from stellar-mass BHs in X-ray binaries and in merging BH binaries, to supermassive BHs in active galactic nuclei (AGNs). A number of astrophysical phenomena are believed to be determined by, or at least associated to it: for instance the launching of relativistic jets; the powering of gamma ray bursts and peculiar supernovae; key properties of accretion disks and hot coronae hovering around BHs. Spin measurements inform also models of BH formation and growth \citep[see, e.g.,][and references therein]{Volonteri2003, Roadmap2019}. Effects originating from BH spin provide diagnostic tools to verify some fundamental predictions of GR and the Kerr hypothesis itself \citep[see e.g.][]{Bambi2011}.

Several methods have been devised to measure or constrain BH spin; they all resort to signals coming from within a few gravitational radii of the BH \citep[see e.g.,][and references therein]{Middleton2016, FalangaBook}. For example, one method exploits the "multi-colour" blackbody X-ray emission from the innermost region of the accretion disk that surrounds stellar mass BHs  
\citep[see e.g.,][and references therein]{McClintock2014}; other methods are based on the 
quasi-periodic oscillations in the X-ray flux of 
accreting BHs and their interpretation in terms of fundamental frequencies of motion of matter orbiting close to BHs \citep[see e.g.,][and references therein]{BelloniStella2014}. The gravitational wave signal detected by LIGO and Virgo from binary BH mergers encodes also information on BH spin \citep[see e.g.,][]{Abbott2019}.

Among spectral methods, the modeling of reflection features in the X-ray spectra of disk accreting BHs of all masses has received a great deal of attention in relation to its potential in measuring BH spin, revealing strong-field GR effects and possibly testing the predictions of alternative gravity theories \citep[see e.g.,][and references therein]{Fabian1989,Laor1991,Reynolds1997,Fabian2000,Martocchia2000,Reynolds2003,Psaltis2008,Johannsen2013, Reynolds2014,Bambi2017b,Zhou2020}. Such features originate in the illumination of the (optically thick) accretion disk by a hot inner Comptonising corona; they include an excess below a few keV, a broad bump starting at a $\sim10$ keV and usually peaking around 30 -- 40 keV, together with the main object of the present study, an iron fluorescence (Fe $K_{\alpha}$) emission line, centred around $\sim6.4$ keV with a remarkably broad, redshifted and asymmetric shape extending over a few keV. The profile of this line is believed to arise from the motion of matter in the inner disk regions as result of a combination of GR effects, relativistic Doppler shifts and beaming, gravitational and transverse redshifts  and  light bending \citep{Fabian1989}. By fitting the observed X-ray spectra with models including the Fe-line profile calculated in the Kerr spacetime by integrating over the line-emitting disk region, it is possible to estimate accretion disk parameters such as the inclination angle, the inner radius, and the radial dependence of the surface emissivity. Most importantly, since the disk is expected to extend down to the innermost stable circular orbit (ISCO), whose radius depends on the spin parameter, the latter can be inferred, or at least constrained \citep[see e.g.,][]{Matt1993,Martocchia1996}: therefore, these profiles represent a diagnostic tool to determine BH spin and to study the accretion flows and motions, as well as the physical conditions of matter in the close vicinity of compact objects, both in X-ray binaries \citep[see e.g.,][and references therein]{Pandel2008,Cackett2010,Miller2010} and AGNs \citep[see e.g.,][]{Tanaka1995,Perez2010,Reynolds2019}. One of the advantages of this method is that no knowledge of BH mass and distance from us is required in order to derive the spin. 

In general axially symmetric spacetimes there exists an infinite number of photon trajectories connecting the emission point to the observer and giving rise to an infinite number of source images. Standard models of relativistic Fe $K_{\alpha}$ line profiles are constructed in the Kerr metric by considering only the first order direct disk image \citep[see e.g.,][]{Dovciak2004,Brenneman2006,Dauser2010}. This is produced by the set of shortest photon trajectories, which do not cross the equatorial plane. Disk higher order images (HOIs) are produced by photon trajectories originating from both sides of the disk and crossing the equatorial plane between the inner edge of the optically thick disk and the event horizon of the BH, the so-called \emph{plunging region} \citep[][]{Reynolds1997,Wilkins2020}. Their contributions to the flux and spectral features from the disk as seen at infinity were explored in previous studies 
\citep[][]{Beckwith2005,Niedzwiecki2018,Bambi2020}.

In this paper we study the contribution of HOIs of co-rotating and counter-rotating disks to the Fe $K_{\alpha}$ line profile in the Kerr spacetime by considering various power-law and lamp-post emissivity profiles for a set of representative values of BH spin and observer inclination angle. These line profiles are then used as input for spectral simulations aimed at determining the conditions under which significant departures from  direct-image-only Fe K$_{\alpha}$ profiles can be singled out through observations of BH  X-ray binary systems with very large area X-ray instruments of next generation, such as those planned for the {\it enhanced X-ray Timing and Polarimetry Mission} (\textit{eXTP} \citealt{Zhang2019}).  
We also discuss the optical depth of the plunging region inside the ISCO and derive an approximate lower limit to it based on the adoption of a different boundary condition at the ISCO radius and disk radiative efficiency.

Our paper is structured as follows: the modelling of the contribution from the HOIs is presented in Sect. \ref{sec:model}; Sect. \ref{sec:simulations} describes the features of our calculated line profiles; simulations are presented in Sect. \ref{sec:data}; a summary and discussion of our results is given in Sect. \ref{sec:summary}.

\section{Modelling the higher order disk images} 
\label{sec:model} 
\subsection{Contribution to the Fe $K_{\alpha}$ lines profiles}
\label{sec:Fe_contirbution} 
We assume a geometrically thin, optically thick equatorial accretion disk around a Kerr BH, with disk matter orbiting in circular geodesic motion down to the ISCO. The spin and mass parameters of the BH are $a$ and $M$. Since we assume a counter-clockwise rotating Kerr BH, the axial vector of its positive spin $a$ points from the BH south pole towards the north pole.
The line element of the Kerr spacetime in Boyer-Lindquist (BL) coordinates ($t, r,\theta,\varphi$) is
\begin{eqnarray}
 \mathrm{d}s^2 &=& g_{\mu\nu}\mathrm{d}x^{\mu}\mathrm{d}x^{\nu}= -\left(1-\frac{2  r}{\Sigma}\right) \,\mathrm{d}t^2 
  - \frac{4 r a }{\Sigma} \sin^2\theta\,\mathrm{d}t\, \mathrm{d}\varphi 
  + \frac{\Sigma}{\Delta}\, \mathrm{d}r^2  \nonumber \\
&+& \Sigma \,\mathrm{d}\theta^2
  + \left(r^2 + a^2 +\frac{2 r a^2 \sin^2\theta}{\Sigma}\right)\sin^2\theta\, \mathrm{d}\varphi^2, 
\label{kerr_metric}
\end{eqnarray}
where 
$\Sigma \equiv r^{2} + a^{2}\cos^{2}\theta$ and $\Delta \equiv r^{2} - 2r + a^{2}$. We adopt units $c = G = M = 1$, so that the gravitational radius $r_g = GM/c^2$ is equal to $1$. The static observer is placed at infinity, $r_{\rm obs}\to\infty$, with angular coordinates ($\phi_{\rm obs}=0$, $\theta_{\rm obs}$). The observer inclination angle 
measured relative to the BH rotational axis is $i \equiv \theta_{\rm obs}$. Photon motion is described by Carter's equations, 
which can be written in the following form \begin{eqnarray}
\label{CarterEQs}
     p^{r} &=& \dot{r} = s_r\Sigma^{-1} \sqrt{R_{\lambda,q}(r)}\,, \nonumber \\
     p^{\theta}& =& \dot{\theta} = s_{\theta}\Sigma^{-1} \sqrt{\Theta_{\lambda,q}(\theta)}\,, \\
     p^{\phi} &=& \dot{\phi} = \Sigma^{-1} \Delta^{-1}
     \left[ 2ar + \lambda \left( \Sigma^{2} - 2r \right)  
\mathrm{cosec}^{2} \theta \right]\,, \nonumber \\
     p^{t}& =& \dot{t} = \Sigma^{-1} \Delta^{-1} \left( \Sigma^{2} -
     2ar \lambda \right)\,, \nonumber
	\end{eqnarray} 
where dotted quantities denote differentiation with respect to the affine parameter, and the sign pair ($s_{r}$, $s_{\theta}$) describes orientation of radial and latitudinal evolution, respectively \citep{Carter1968}.
The radial and latitudinal effective potentials are
\begin{eqnarray}
   \label{eqn:2.1.3}
       R_{\lambda,q} \left( r \right) &=& \left[ \left( r^{2} + a^{2}
\right) - a \lambda \right] ^{2}
       - \Delta \left[ q + \left( \lambda - a \right) ^{2} \right]\,, \nonumber \\  
       \Theta_{\lambda,q} \left( \theta \right) &=& q + a^{2} \cos^{2}
\theta -\lambda^{2} \mathrm{cot}^{2} \theta\,. 
\end{eqnarray} 
Here $\lambda$ and $q$ are constants of motion related to the photons covariant angular and linear momenta. In the case of a distant observer the rays reaching the observer position are virtually parallel and the relations between Cartesian coordinates on the detector screen and the constants of motion can be written as \citep[][]{Cunningham1973}
\begin{equation}
x=-\frac{\lambda}{\sin\,\theta_{obs}}\,,  \qquad y=\Theta_{\lambda,q} \left( \theta_{obs} \right)\,.
\end{equation}

\begin{figure}
\centerline{\psfig{figure=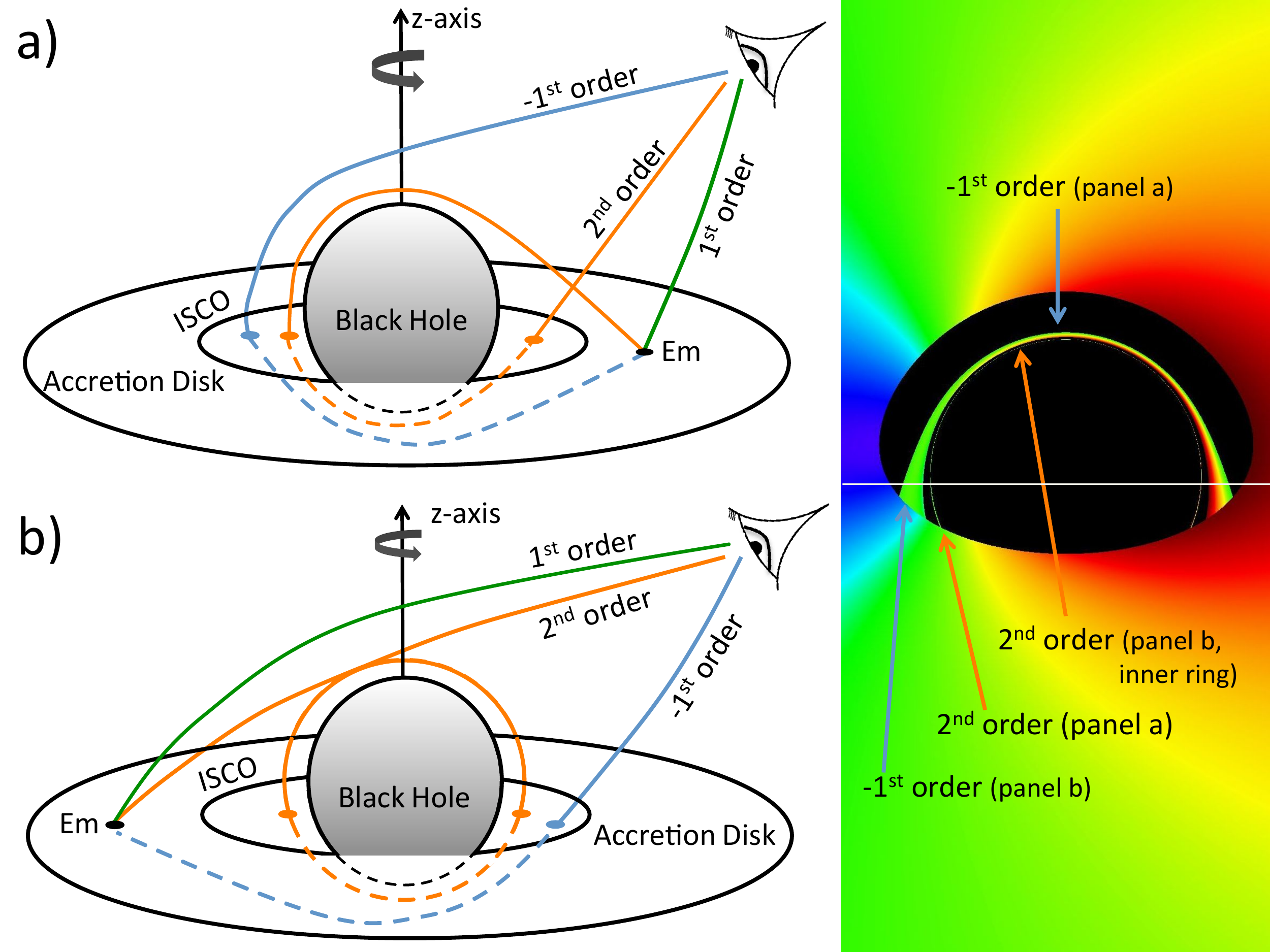, scale=0.32}}
\caption{Illustrative sample of photon trajectories emitted at the point $E_{\rm m}$ having spherical coordinates (r$_{\rm em}$,$\pi/2$, $\varphi_{\rm em}$) from an accretion disk around a BH. The first order trajectory produces a direct image. The minus-first order trajectory originates from below the disk, passes between the BH and the ISCO and gives rise to the first indirect image in the the observer's plane. The trajectory looping more than $2\pi$ around the BH gives rise a second order direct image. The colors in the right panel represent the frequency shift factor, $g$, of the radiation coming from the disk surface and giving rise to different order images. The horizontal line corresponds to the cross-section between the observer local equatorial and BL coordinate equatorial plane.}
\label{fig:fig1}
\end{figure} 

\begin{figure}
\centering
\hbox{
\psfig{figure=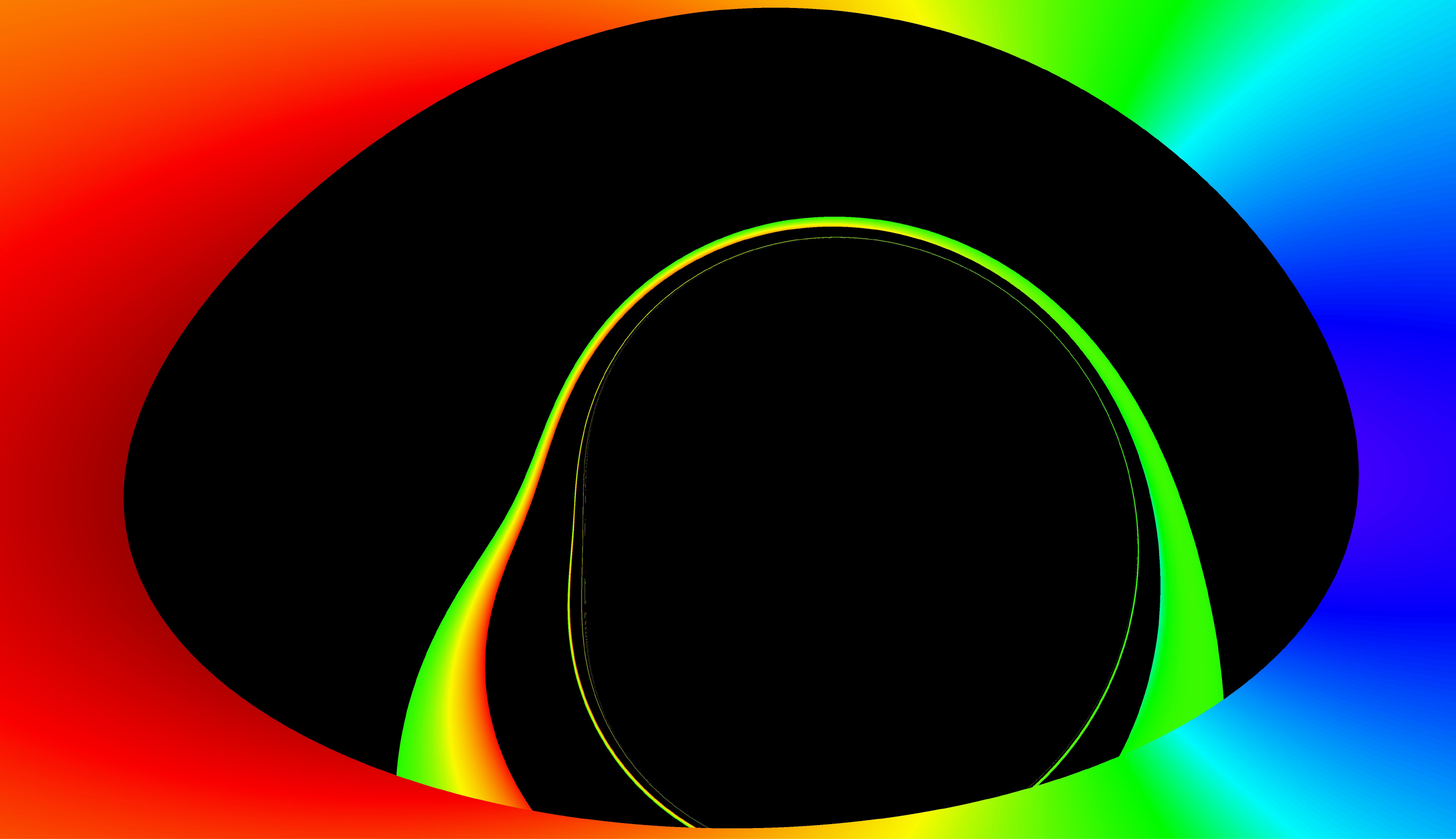, angle=-90, scale=0.28}
}
\vspace{0.1cm}
\hbox{
\psfig{figure=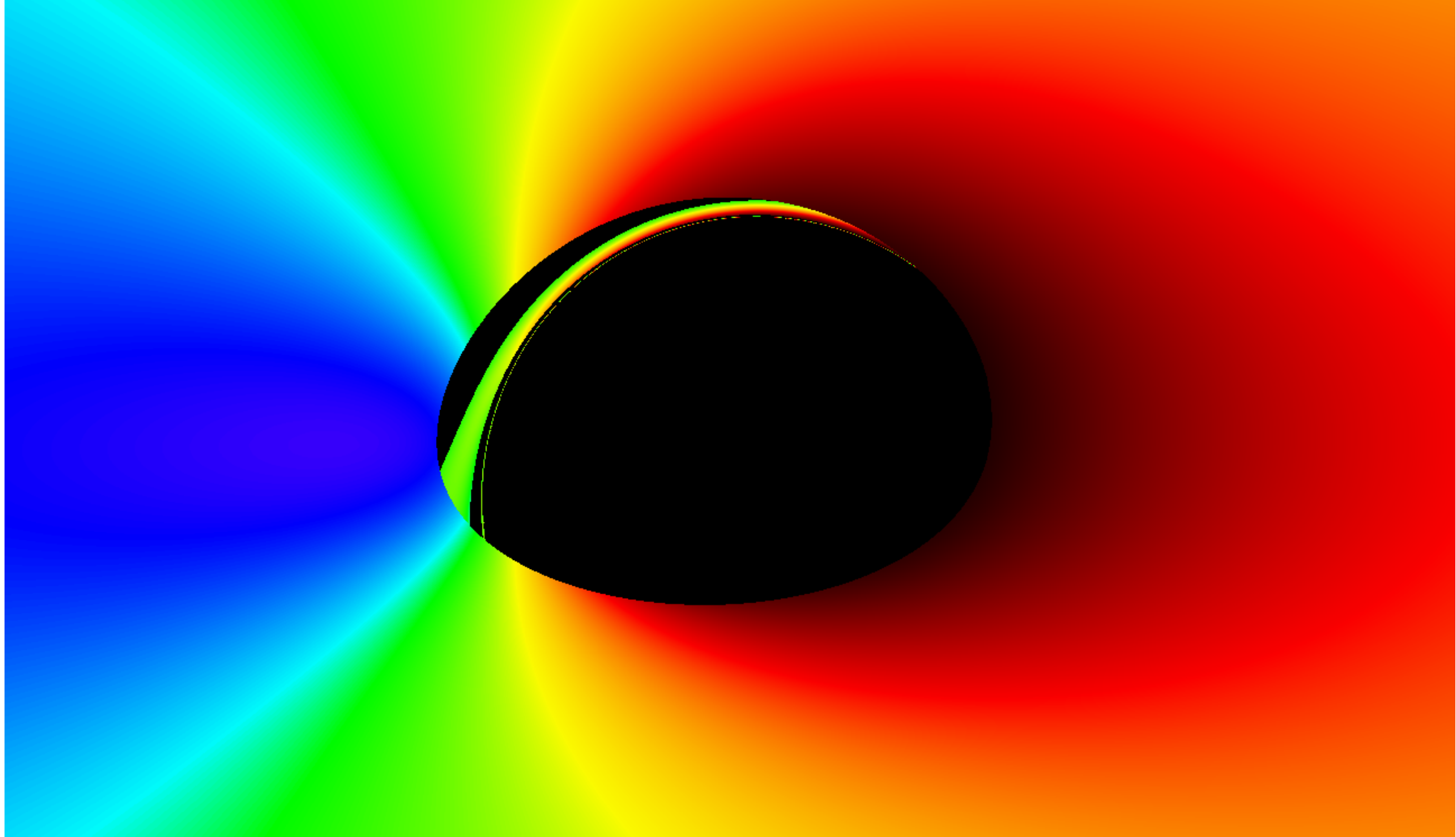, angle=-90, scale=0.364}
}
\caption{ A visualization of an inner part of the accretion disk around a BH showing also disk HOIs inside the BH shadow. Upper panel: counter-rotating disk with $a = -0.9995$ and the disk inner edge at $R_{\rm ISCO\, -} \simeq 8.99\,r_g$, the most extreme case considered here. Lower panel: co-rotating disk with $a = 0.6$ and the disk inner edge at $R_{\rm ISCO\,+} \simeq 3.83\,r_g$, a case discussed in Sect.4 . Both panels are identically zoomed with respect in the observer's sky and displayed for the observer inclination angle $i=60^{\circ}$. The colour scale represents the frequency shift factor, $g$, of the emission, the green colour corresponding to unity.}
\label{fig:fig2}
\end{figure}

Owing to the extreme lensing properties of BHs, photons from the accretion disk can reach the observer along an infinite number of trajectories. Photons coming from the upper disk side (that facing the observer) give rise to {\it direct} images. Among these, photons travelling along the shortest trajectories produce the usual, {\it first order direct} image. {\it Second and higher order images} arise from photon trajectories undergoing one or more loops in the close vicinity of the BH. The $n$-th order {\it direct} image corresponds to photons crossing the equatorial plane $2(n-1)$ times. Photons that reach the observer being emitted from the lower disk side give rise to $n$-th order {\it indirect} images crossing the equatorial plane $2n-1$ times \citep[see, e.g.,][]{Viergutz1993,Beckwith2005}. Figure \ref{fig:fig1} illustrates selected photon trajectories, with a negative sign marking those corresponding to indirect images. Hereafter, all images except the first direct image are referred to as HOIs. 

On the observer's screen, the first direct photons produce the primary main image of the disk around the BH, whereas the HOIs are located within the BH shadow (see Fig. \ref{fig:fig1} and \ref{fig:fig2}). In the case of a geometrically thin, optically thick disk extending down to the ISCO, the boundary of the so-called BH shadow is given by the ISCO image as seen from an inclination angle $i$ \citep[see e.g.][and references therein]{Luminet1979,Falcke2017,Dokuchaev2019}. The radial coordinate of the ISCO in the Kerr spacetime is 
\begin{equation}
\label{RISCO}
R_{\rm ISCO\, \pm}=3+Z_2\mp\sqrt{(3-Z_1)(3+Z_1+2Z_2)}\,,
\end{equation}
where $Z_1=1+(1-a^2)^{1/3}((1+a)^{1/3}+(1-a)^{1/3})$ and $Z_2=(3a^2+Z_1^2)^{1/2}$. The subscripts $+$ and $-$ denote the case of a co-rotating and counter-rotating motion of the disk matter relative to the BH spin, respectively. In the non-rotating (Schwarzschild) BH case with $a=0$ the ISCO is located at $R_{\rm ISCO}=6~\mathrm{r_g}$. The radial coordinate of the co-rotating ISCO approaches the outer event horizon for increasing values of $a$, and, in the case of extreme spin $ a = 1 $, it merges with the horizon at $R_{\rm ISCO+}=1~\mathrm{r_g}$. Instead, the radius of the counter-rotating ISCO  moves away for increasing spins, reaching its extreme at  $R_{\rm ISCO-}=9~\mathrm{r_g}$.

The closest approach of photons reaching the observer to the BH is in all cases limited by the radius of the equatorial co-rotating circular photon orbits (CPO+), as all radial turning points of photons escaping to infinity are located above it \citep[see e.g.,][]{Teo2003}. The radial coordinate of the CPO+ is 
\begin{equation}
r_{\mathrm{cpo}+} = 2 \left[1+\cos \left(\frac{2}{3} \arccos \left(-a\right)\right)\right]\,.
\end{equation}
The CPO+ is at $r=3~\mathrm{r_g}$ in the Schwarzschild case ($a=0$) and drifts towards the event horizon for increasing spin. In the extreme Kerr case ($a=1$), the CPO+ touches the event horizon at $r=1~\mathrm{r_g}$. The image of the CPO+ on the observer screen determines the minimum angular extent that HOIs can reach \citep[see e.g.,][for details]{Cunningham1973,Viergutz1993,Dokuchaev2019}.

The frequency shift factor $g$ (the ratio of the emitted and observed photon energy) due to the circular orbital motion of the disk matter in the equatorial plane ($\theta=\pi/2$), can be expressed using the metric coefficients (see Eq. \ref{kerr_metric}) as \citep{Bardeen1972}
\begin{equation}
g=\frac{\sqrt{-g_{t t}-2 g_{t \phi} \Omega_K-g_{\phi \phi} \Omega_K^{2}}}{1-\lambda \Omega_K}\,, 
\end{equation}
where 
\begin{equation}
\label{Komega}
\Omega_{K \pm}=\frac{\pm 1}{r^{3 / 2} \pm a } 
\end{equation}
is the Keplerian angular velocity in Kerr metric. The observed flux 
at the energy $E_c$ per solid angle, $\Delta \Pi$, subtended by the disk on the observer screen is given as \citep{Misner1973},
\begin{equation}
\label{ebin_flux}
\Phi(E_c) =  {\int_{\Delta \Pi}}I_{\rm em}(r,\alpha_{\rm em})g^{3}\,f(E_c/g,\sigma,E_0) \,  {\rm d}\omega\,,
\end{equation}
where $I_{\rm em}(r,\alpha_{\rm em})$ is the local emissivity generally depending on the radius $r$ and the local photon emission angle $\alpha_{\rm em}$, d$\omega$ is the solid angle element and we approximate the very narrow double-peak rest profile of the Fe $K_\alpha$ line by a Gaussian function $f(E_c/g,\sigma,E_0)$ centred at $E_0=6.4$ keV with FWHM $\sigma=5$ eV \citep{Basko1978,Bakala2015}. Of course the flux equation (\ref{ebin_flux}) can be used with any disk spectral component, as represented by the appropriate spectral distribution function $f(E_c/g)$.

\subsection{Optical Depth of the Plunging region}
\label{sec:Plunging_region} 
\begin{figure}
\begin{center}
\vbox{
\includegraphics[trim=1cm 0cm 0cm 
0cm,scale=0.3]{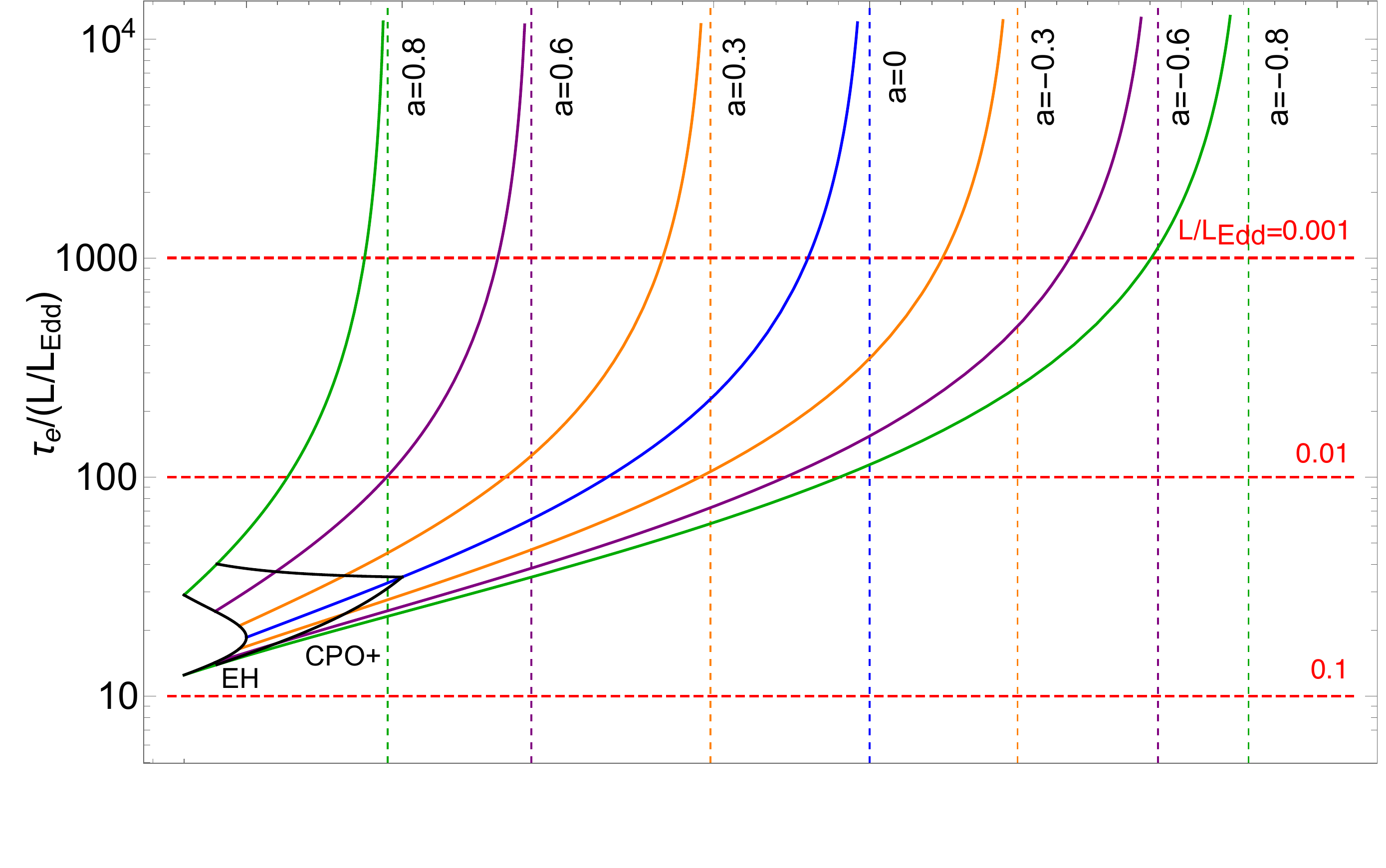}\\
\vspace{-0.6cm}
\includegraphics[trim=1cm 0cm 0cm 0cm,scale=0.3]{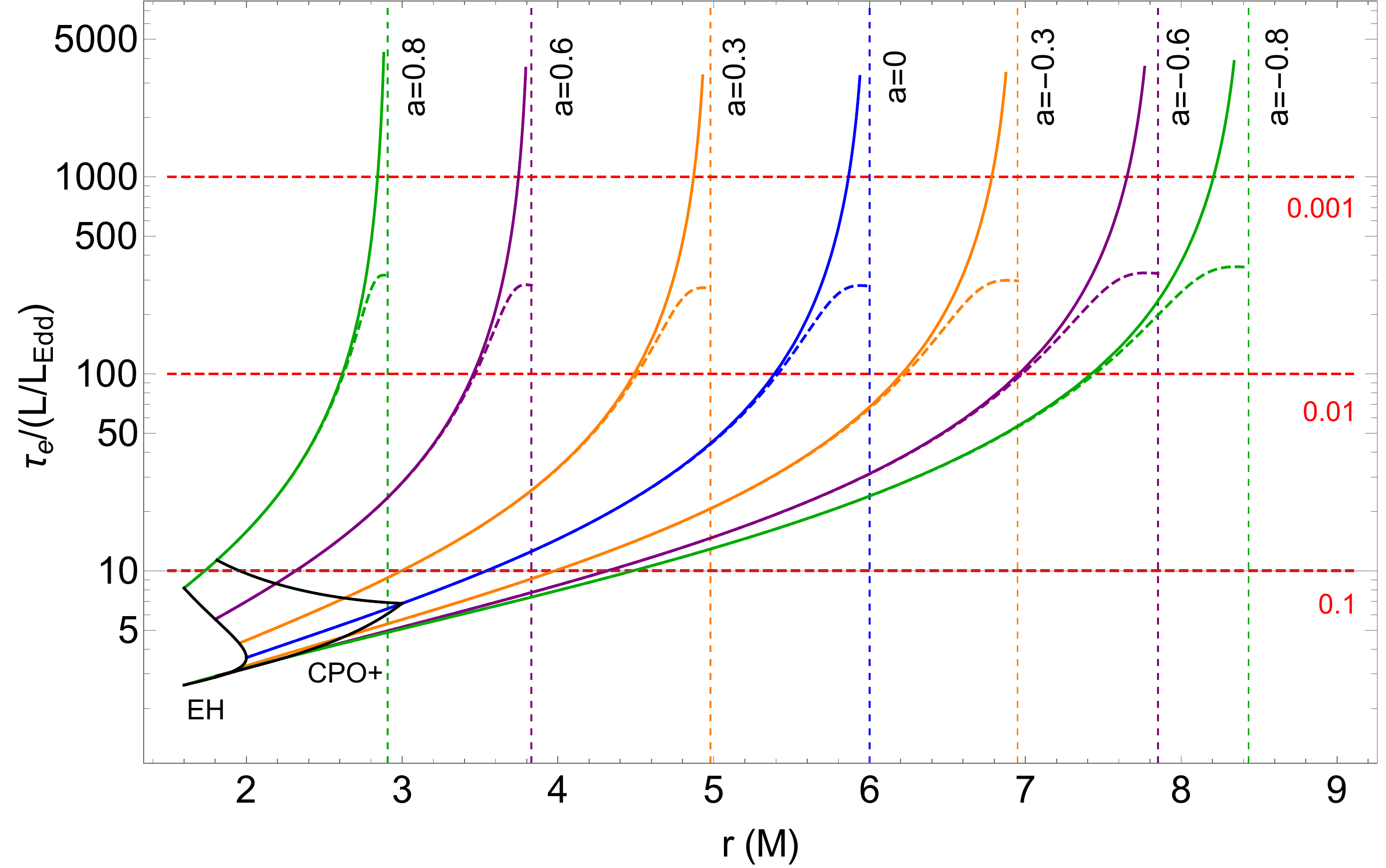}
}
\end{center}
\caption{Optical depth of the plunging region as a function of radius for selected values of the spin parameter $a$: \textit{top} - profiles for standard disk radiative efficiency and null radial velocity boundary condition at the ISCO; \textit{bottom} - profiles for the case with the modified disk radiative efficiency given by Eq. (\ref{eq:etaim}) and null- and non-null radial velocity (using $\mathcal{A}=0.01$) boundary condition (dashed and continuous curves in the lower panel, respectively). 
The vertical dashed lines represent the ISCO radius and the dashed horizontal red lines are set at $\tau_e=1$ for luminosities $L/L_{\rm Edd}=0.1,0.01,0.001$, which delimit the optically thick (upper part) and thin (lower part) regimes (note the luminosity scaling
of the of the Y-axis units). The profiles are limited from below by the innermost black curve, representing the horizon radius; the black curve at slightly larger radii represents the CPO+ radius, giving a lower limit for equatorial plane crossing points of HOI rays.}
\label{fig:FigTot}
\end{figure}

In crossing the equatorial plane HOI rays may be attenuated or even obscured by the presence of disk matter spiralling inside the ISCO in the plunging region. The electron scattering optical depth $\tau_e$ in the plunging region is obtained by imposing the conservation of rest mass in the flow and expressed as a function of radius by \citep{Reynolds1997,Wilkins2020}, 
\begin{equation} \label{eq:tauR}
\tau_e=\frac{2}{r \eta(-u^r)}\frac{L}{L_{\rm Edd}},    
\end{equation}
where $L$ is the source luminosity as measured by an observer at infinity, $L_{\rm Edd}$ is the Eddington luminosity, $\eta$ is the mass to radiation conversion efficiency ($L= \eta \dot M$ with $\dot M$ the mass accretion rate) and $u^r(r)$ is the radial geodesic velocity of the inspiralling matter, whose expression in Kerr spacetime is \citep{Chandrasekhar1950}
\begin{equation}
\label{rvelocity}
u^r=-\frac{\left[\left(E\left(r^{2} + a^{2}
\right) - aL_z\right)^{2}
       - \Delta \left( r^2 + \left( L_z - aE \right) ^{2} \right)\right]^{1/2}}{r^2}\,.
\end{equation} 
The motion is determined by the two conserved quantities, the specific energy $E$ and the specific angular momentum $L_{\rm z}$, in the motion from the ISCO through the plunging region; these can be expressed as 
\begin{eqnarray}
E \equiv E_{\rm K}&=&\frac{r^{3/2}-2r^{1/2}\pm a}{r^{3/4}(r^{3/2}-3r^{1/2}\pm 2a)^{1/2}}\Bigg|_{r=R_{\rm ISCO\pm }(a)},\label{eq:enpart}\\
L_{\rm z}\equiv L_{\rm K}&=&\frac{\pm(r^{2}\mp 2ar^{1/2}+ a^2)}{r^{3/4}(r^{3/2}-3r^{1/2}\pm 2a)^{1/2}}\Bigg|_{r=R_{\rm ISCO\pm }(a)}.\label{eq:ampart}
\end{eqnarray}

Standard disk models assume that torquing by viscosity ceases at the ISCO radius, as freely inspiralling motion ensues inside of it \citep[see e.g.,][]{ShakuraSunyaev1973,NovikovThorne1973}. As a result the (inward) radial velocity of disk matter vanishes at the ISCO ($u^r(r_{\rm ISCO\pm})=0$) and the density diverges. By solving Eq. (\ref{rvelocity}) with the above boundary conditions at the ISCO, the optical depth profile 
is calculated from Eq. (\ref{eq:tauR}) for different spin values and the corresponding efficiencies $\eta = 1-E_K $ (see Fig. \ref{fig:FigTot}).

Modern accretion disk simulations indicate that magneto-hydro-dynamic turbulence is the primary torque agent and that magnetic stress may operate well inside the ISCO  \citep[see e.g.,][]{Hawley01,Armitage01}. In application to Eq. (\ref{eq:tauR}), this has two consequences: $u^r(r_{\rm ISCO\pm})$ is no longer zero and the efficiency $\eta$ increases as a result of angular momentum extraction and energy dissipation from inside the ISCO. Simple estimates of the impact of these two effects on the optical depth of the plunging regions are given in the following. We express the radial velocity at the ISCO as a fraction $\mathcal{A}$ of the local Keplerian velocity, {\it i.e} $v^r = \mathcal{A}\, v^\phi $. Considering Eq. (\ref{rvelocity}), the new boundary condition is
\begin{equation} \label{eq:BC}
u^r(R_{\rm ISCO\pm })=v^r(R_{\rm ISCO\pm }).
\end{equation}
This entails modifying the calculation of the geodesic velocity $u^r$, by first fixing the specific angular momentum to the Keplerian value (i.e., $L_z=L_{\rm K}$), and then determining the matter specific energy $E_{\rm n}$ from the boundary condition (\ref{eq:BC}); we obtain
\begin{equation}
E_{\rm n}=\frac{\pm 2 a L_{\rm K}+\sqrt{4a^2L_{\rm K}^2-\rho F }}{\rho}\Bigg|_{r=r_{\rm ISCO\pm}},\\
\end{equation}
where $F=r^3v_r^2-L_{\rm K}^2(r-2)-\rho+2 (r^2+a^2)$ and $\rho=r^3+a^2r+2a^2$.
We provide here also an estimate of disk radiative efficiency $\eta$ that can result from matter torquing inside ISCO; this in turn converts to lower values of the optical depth of the plunging zone. \citet{Agol2000} determined the conditions under which BH spin equilibrium ensues as a result of matter accretion from an increased-efficiency disk extending inside the ISCO. In such a disk the 
angular momentum transferred from inside the ISCO outwards causes a lower spin-up torque by the accreting matter, whereas the increased emissivity in the innermost disk regions leads to a higher spin-down torque by photons captured by the BH. The equilibrium condition is expressed in terms of the disk spin-equilibrium efficiency through the approximate formula 
\begin{equation} \label{eq:etaim}
\eta=1-\frac{\sqrt{C}}{2-B}\Bigg|_{r=R_{\rm ISCO\pm }(a)},    
\end{equation}
where $B=1 \pm a/r^{3/2}$ and  $C=1-3/r \pm 2a/r^{3/2}$ \citep[see][for more details]{Agol2000}.
Accordingly the radiative efficiency for $a=-0.8$, 0, and 0.8 is $\eta=0.19$, 0.29  and 0.43 (here and elsewhere negative values of the spin $a$ refer to counter-rotating disks), respectively, to be compared with the standard \citet{NovikovThorne1973} values $\eta=0.04$, 0.06 and 0.12. By adopting the above prescriptions we calculate what may be regarded as an approximate lower limit to the radial dependence of the optical depth in the plunging region. This is shown by the curves in the bottom panel of Fig. \ref{fig:FigTot}, which can be compared with the corresponding curves in the upper panel, based on the standard treatment for selected values of the BH spin parameter \citep{Reynolds1997}. All curves are plotted for a luminosity of $L = L_{\rm Edd}$ and can be easily rescaled for other $L/L_{\rm Edd}$ values (note the Y-axis definition). In all cases the optical depth is highest at the ISCO and decreases rapidly towards the event horizon. It is seen that the finite radial velocity at the ISCO removes the divergent behaviour and gives rise to lower values and a flatter radial dependence of the optical depth for a small range of radii close to the ISCO. For smaller radii than $\sim 0.9\ r_{\rm ISCO}$ the behaviour becomes virtually identical to that of the solution with $u^r(r_{\rm ISCO\pm})=0$ owing to the very rapid inspiralling of matter towards the horizon. The spin-equilibrium radiative efficiency induces instead an overall downshift of the optical depth curve in accordance with Eq. (\ref{eq:etaim}). Figure \ref{fig:FigTot} shows also the radius of the event horizon and that of the CPO+. Most strongly deflected HOI trajectories cross the plunging zone at radii slightly larger than the CPO+ radius, a region characterised by $\tau_e < 1$ for all curves, as long as $L\lesssim 0.01~L_{\rm Edd}$. In the Fe-line profiles and simulations calculated in the following we assume that the optical depth is negligible throughout the entire plunging region and that the disk remains optically thick and geometrically thin, without developing an inner radiatively-inefficient inner region \citep{Narayan1995,Abramowicz1995}.

\section{Calculation of higher order disk images and line profiles} 
\label{sec:simulations} 

\subsection{The case of power-law emissivity profile}
\label{sec:disksimulations} 
We use our general relativistic ray-tracing code, \emph{LSDplus}, which carries out time-reversed integration of Eqs. (\ref{CarterEQs}) for all null geodesics reaching the observer screen at infinity \citep[see][for more details]{Bakala2015}. The screen resolution is set to $7500 \times 7500$ pixels. The first order direct image of the inner edge of the disk, which we assume to coincide with the ISCO, delimits the BH shadow in our case. The outer edge of the disk is located at $900$ r$_g$.
Following the photon paths from the observer screen to the emission point, the code counts the number of times photon trajectories cross the equatorial plane at a distance at a radial coordinate between the ISCO and the CPO+. This allows singling out the contribution from the HOIs to the images in the observer screen as well as the line profiles (see Figs \ref{fig:fig1} and \ref{fig:fig2}). 
The highest order disk images that our ray-tracing code can process is limited only by screen resolution; in practice the flux contributions from third and HOIs can be considered negligible \citep[see e.g.,][]{Ohanian1987}. The Fe $K_{\alpha}$ line profile is calculated by summing the flux from each pixel of the observer screen in the relevant energy bin of the profile, see Eq. (\ref{ebin_flux}). The radial dependence $\mathcal{R}(r)$ of the disk line emissivity is generally assumed to be a power-law function of the radial coordinate; we also adopt an isotropic dependence of the emissivity on the local emission angle $\alpha_{em}$, thus: 
$I_{em}(r,\alpha_{em}) = \mathcal{R}(r) \propto r^{q}$, where we use the classical prescription for the radial dependence of the disk emissivity, $q = -3$. Except for the innermost disk regions, this approximates well the radial emissivity in the so-called {\it lamp-post} geometry (see Sec. \ref{sec:lampsimulations} for details).

\begin{figure*}
\centering
\hbox{
\psfig{figure=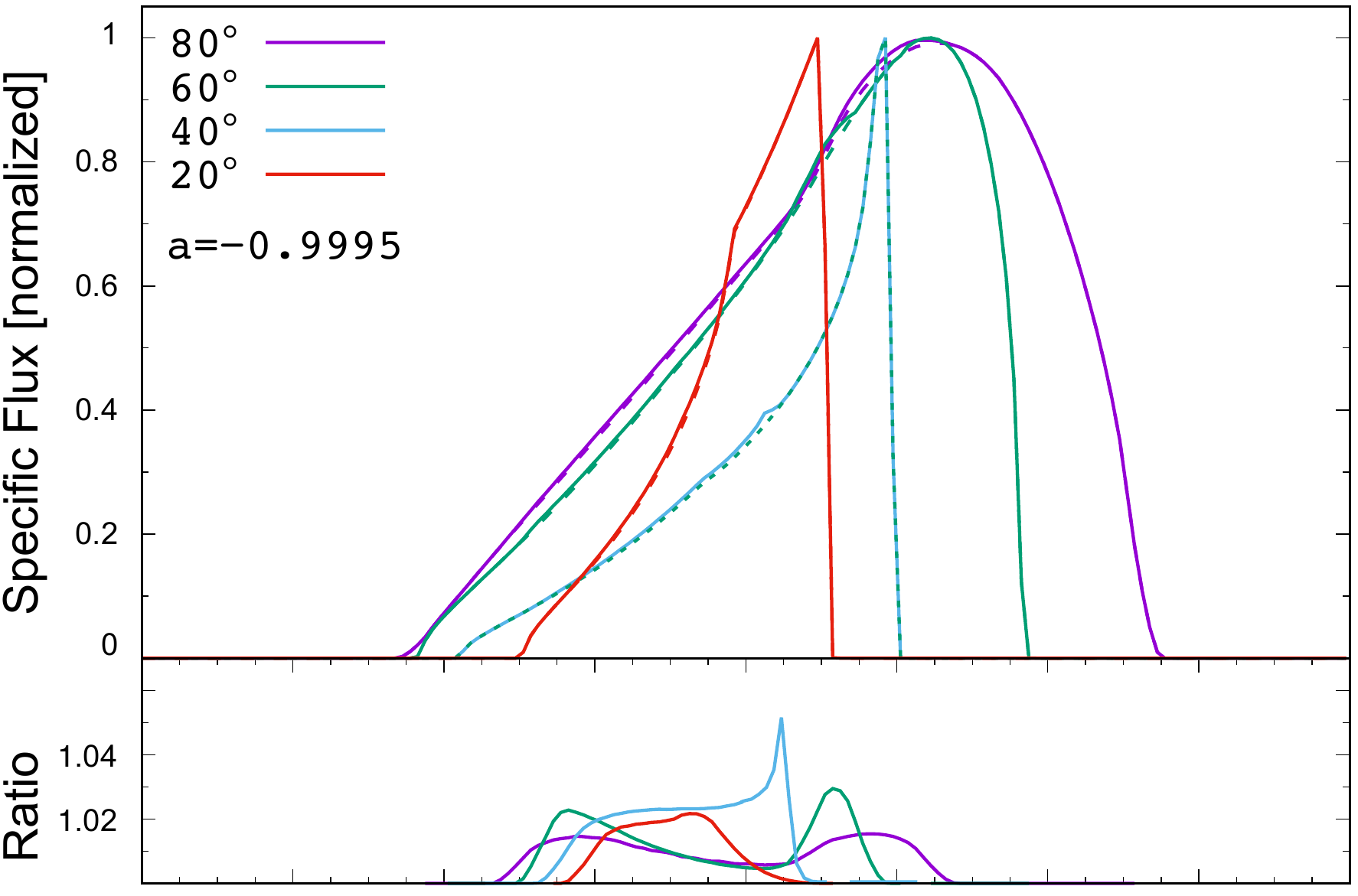, scale=0.35}
\hspace{-0.25cm}
\psfig{figure=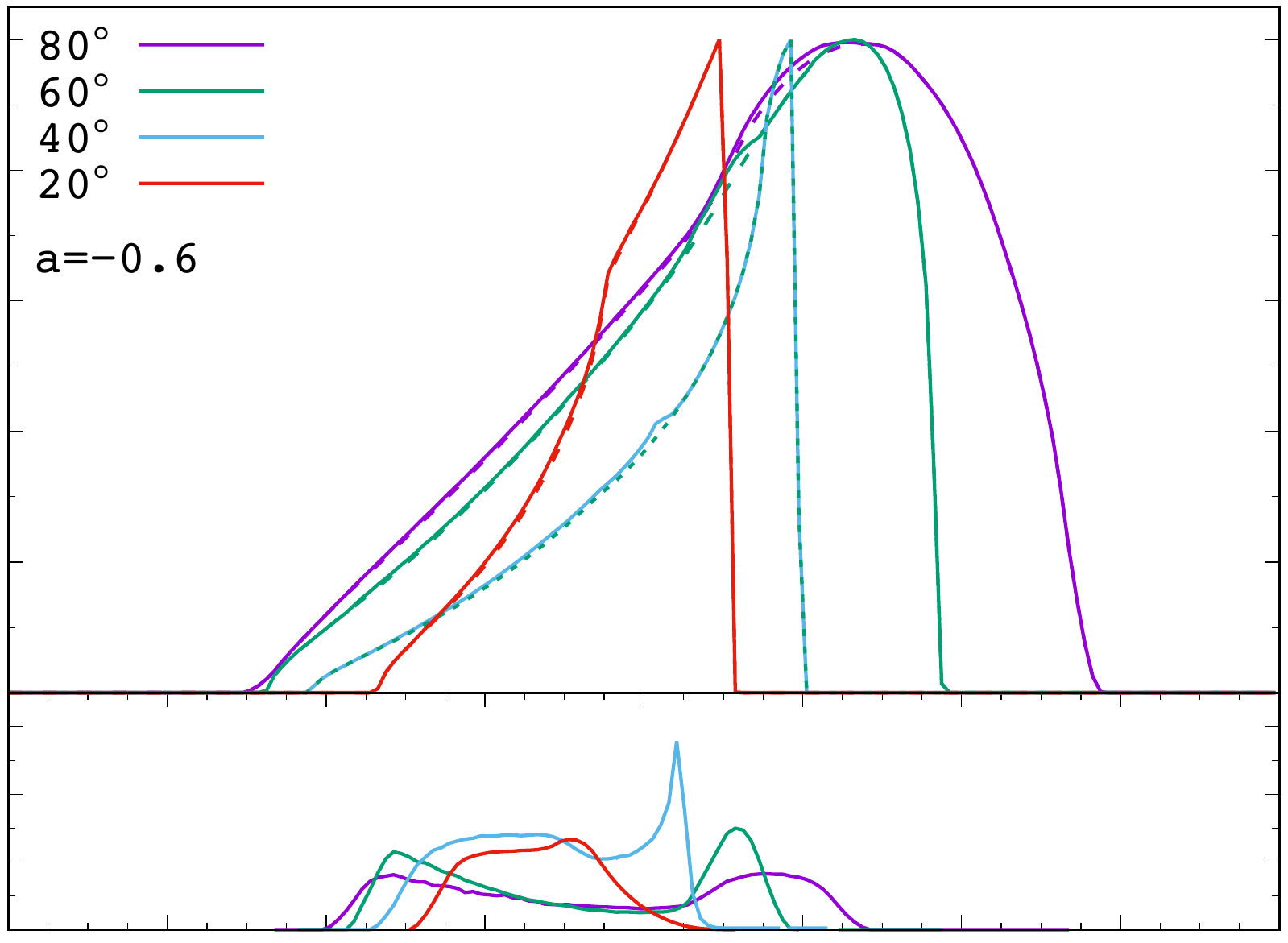, scale=0.35}
\hspace{-0.25cm}
\psfig{figure=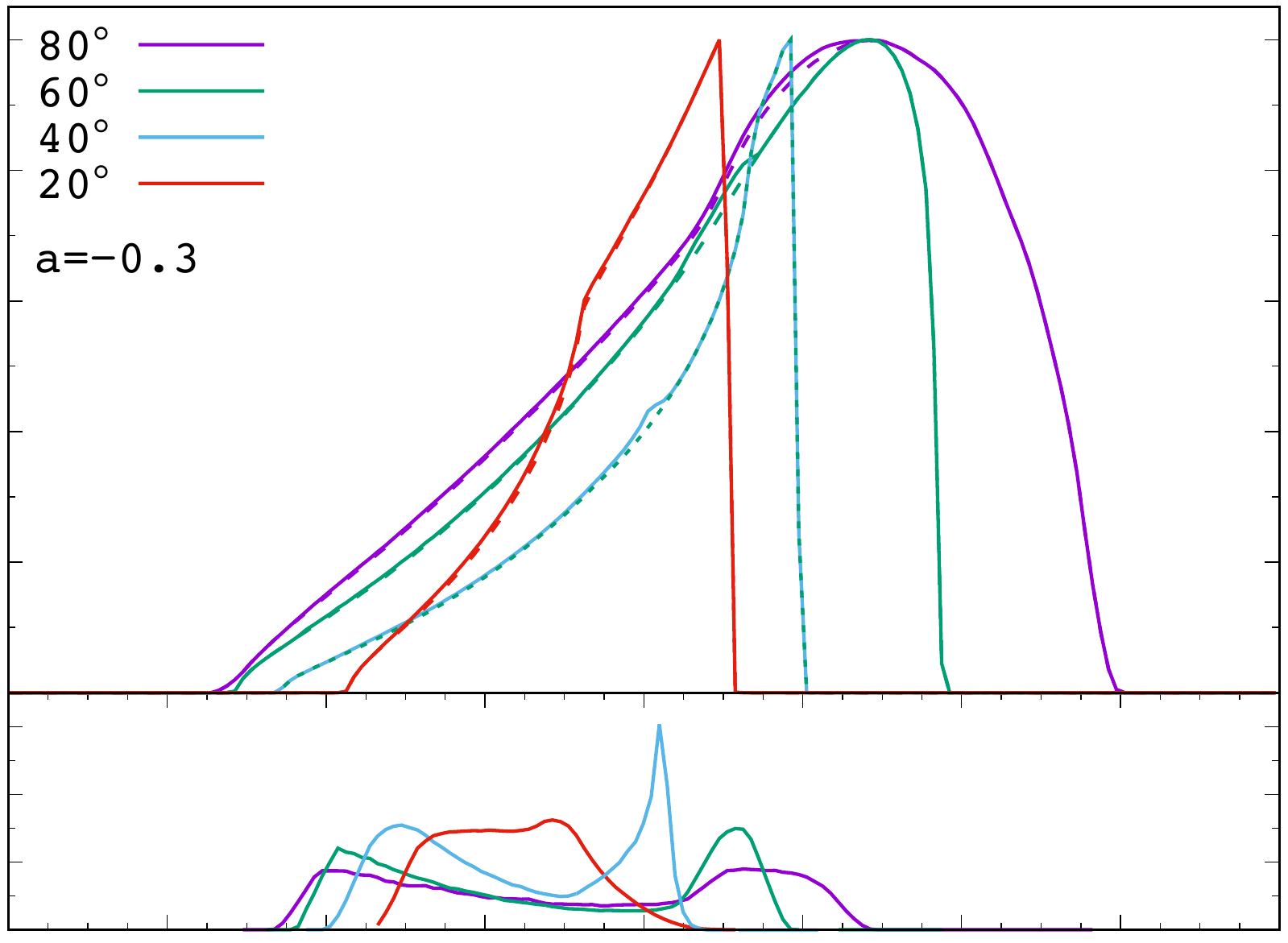, scale=0.35}
}
\hbox{
\psfig{figure=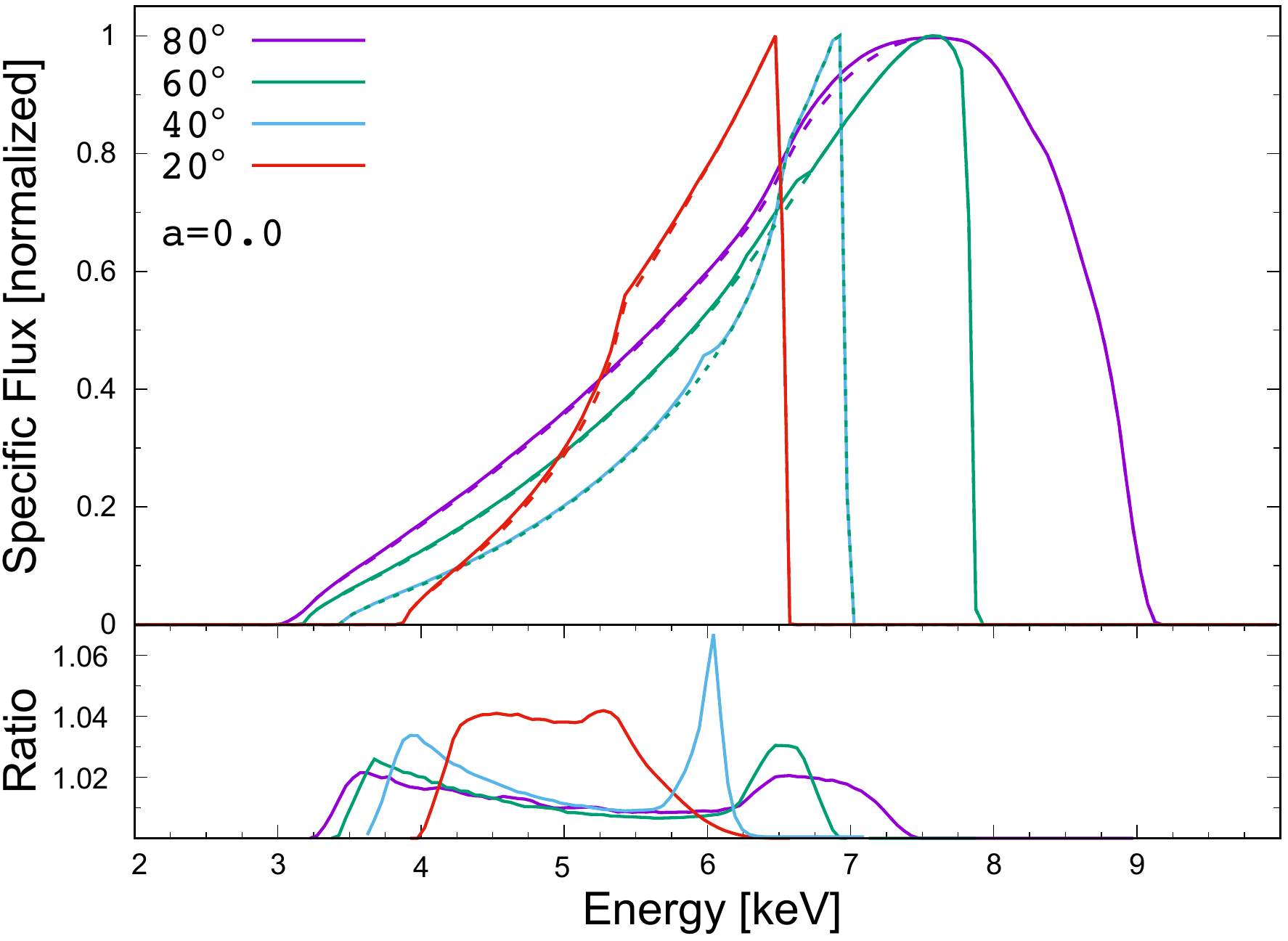, scale=0.35}
\hspace{-0.25cm}
\psfig{figure=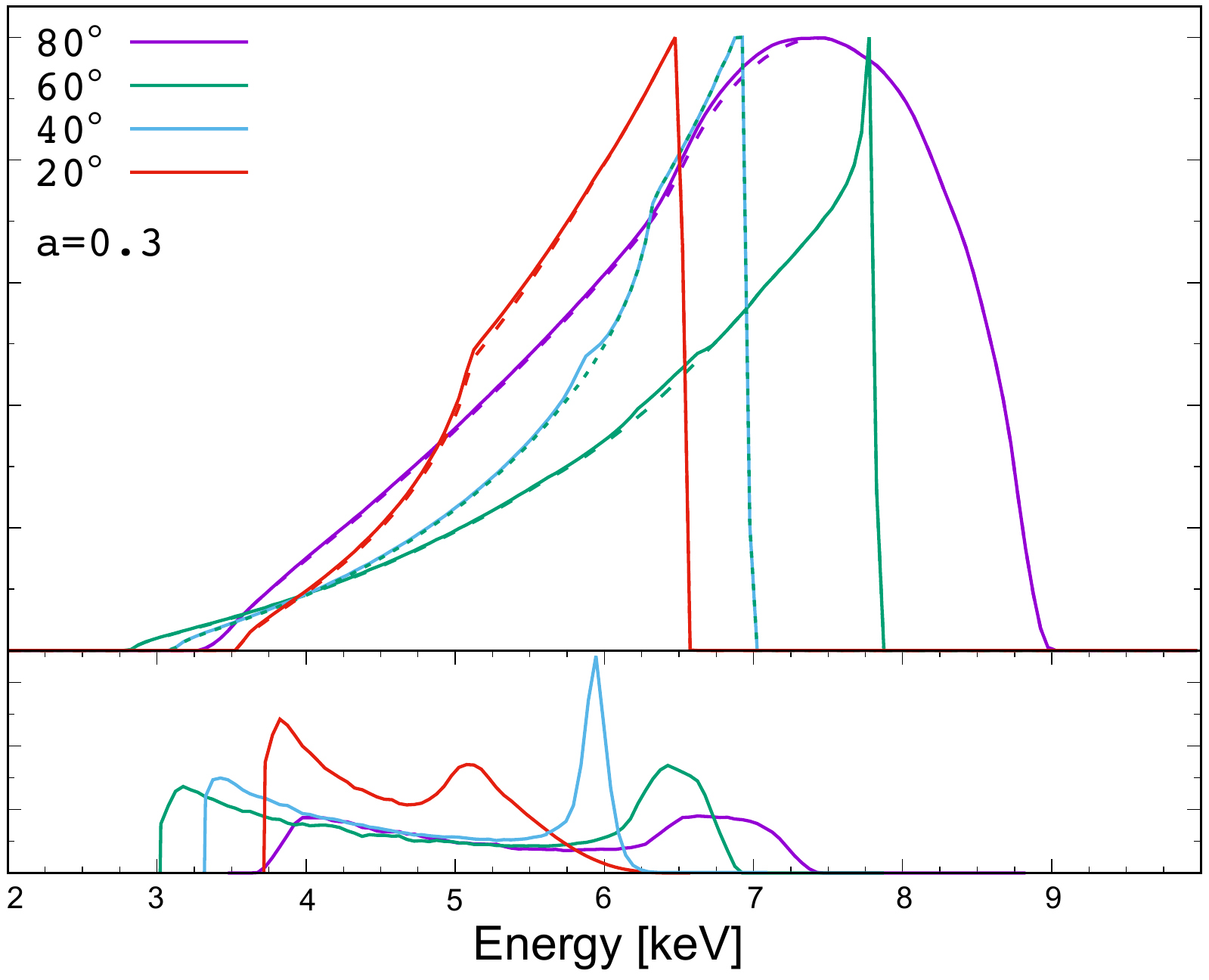, scale=0.35}
\hspace{-0.25cm}
\psfig{figure=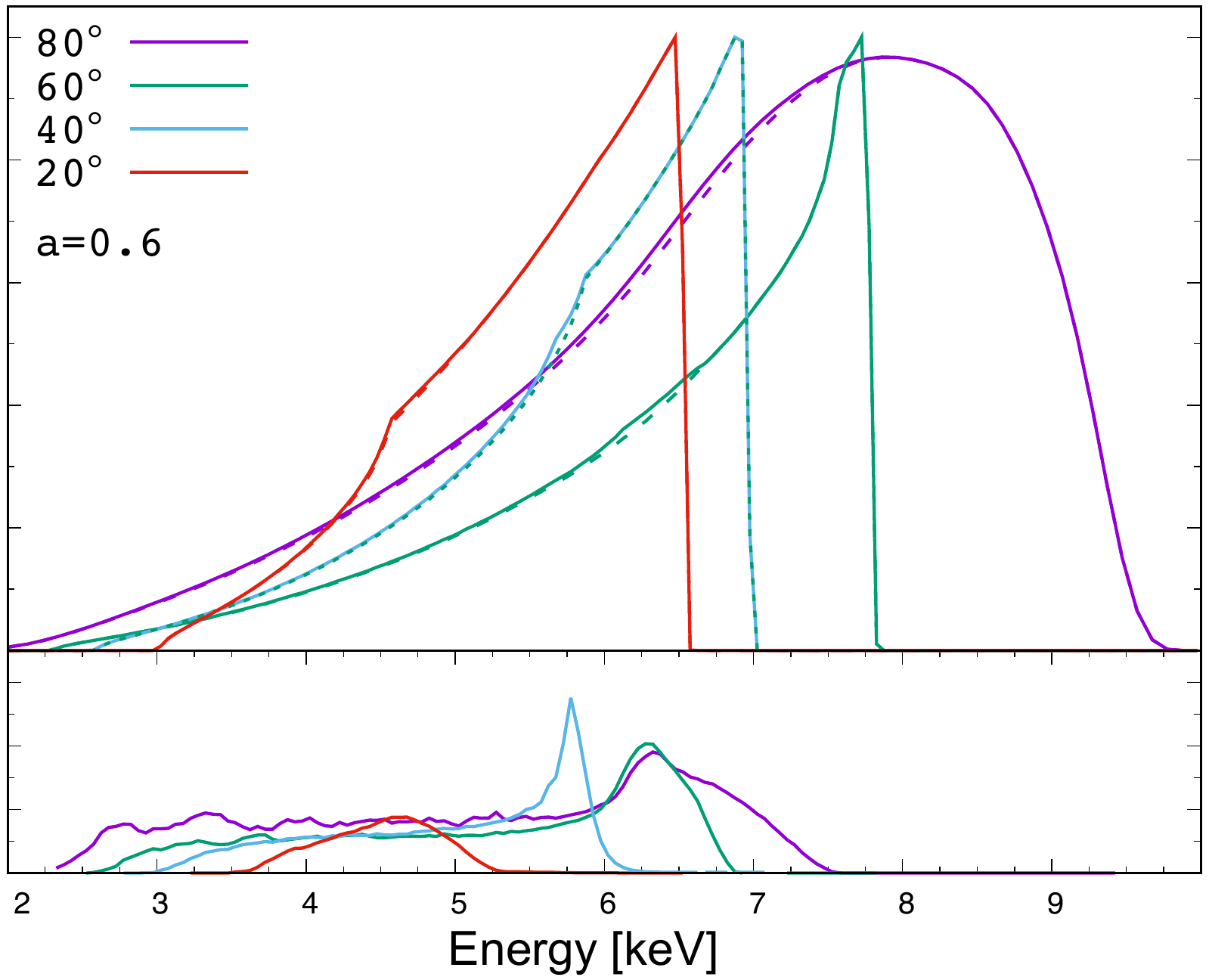, scale=0.35}
}
\caption{
Selected Fe $K_{\alpha}$ line profiles with (solid lines) and without (dashed lines) disk HOIs contributions for different BH spin values, $a$. The colors represents different observer inclination angles. The bottom panels display the ratio between specific fluxes with and without high order images contributions.}
\label{fig:Fig4}
\end{figure*}

The panels in Fig. \ref{fig:Fig4} show our calculated line profiles for different inclination angles ($i=20^{\circ}$, $40^{\circ}$, $60^{\circ}$, $80^{\circ}$) and BH spins ($a=0$, $\pm0.3$, $\pm0.6$, $-0.9995$). All profiles are normalized to their peak specific photon flux. The characteristic HOI features imprinted in the profiles can be singled out through comparison with the corresponding profiles from which the contributions of the HOIs have been excluded (see the dashed curves). 
The bottom sub-panels show the ratio of each pair of profiles.  
For increasing spin values of co-rotating disks, HOIs are progressively occulted as the ISCO approaches the event horizon; they are blocked entirely for $a=1$. For this reason, we do not show the line profiles for a co-rotating disk with nearly extreme spin, $a=0.9995$. The contribution from the HOIs is driven mainly by: $i$) the spin-dependent width of the region between the ISCO and CPO+, where HOI photons cross the equatorial plane; $ii$) the observer inclination angle, which determines the HOI observed energy range, as well as the partial occultation of the HOIs by the inner disk rim closest to the observer itself; $iii$) the angular size of HOIs, which decreases for increasing spins, as the CPO+ gets progressively closer to the BH event horizon. 

From Fig. \ref{fig:Fig4} it is apparent that the broad peak on the blue side of the profiles remains virtually unaffected by HOIs. In general, the contribution of the HOIs tends to be most pronounced in a wide energy interval placed around the center of the profile and extending close to the end of the red wing. Specific flux ratios in this interval have typical values in the $\sim $1\% -- 4\% range.  
The highest peaks in ratio ($\sim$5\% -- 7\%) are attained for $i=40$\textdegree; slightly decreasing values of the ratio are found for increasing spins. In the profiles that we studied overall contribution of the HOIs to the Fe-line flux ranges between $\sim 0.3$ and $1.4$\%. For lower inclinations, $i=20^{\circ}$, HOI occultation by the inner disk rim is totally absent. Increasing absolute values of the spin $a$ and the resulting shrinkage of the CPO+ and solid angles subtended by the disk HOIs lead to a decreasing angular size of the HOIs and a reduced contribution of the HOIs to the line profiles. This can be seen in Fig. \ref{fig:Fig4} for $a=0.6$ and especially for $a=-0.6$ and $-0.9995$, owing to the larger ISCO size.

\begin{figure*}
\hbox{
\psfig{figure=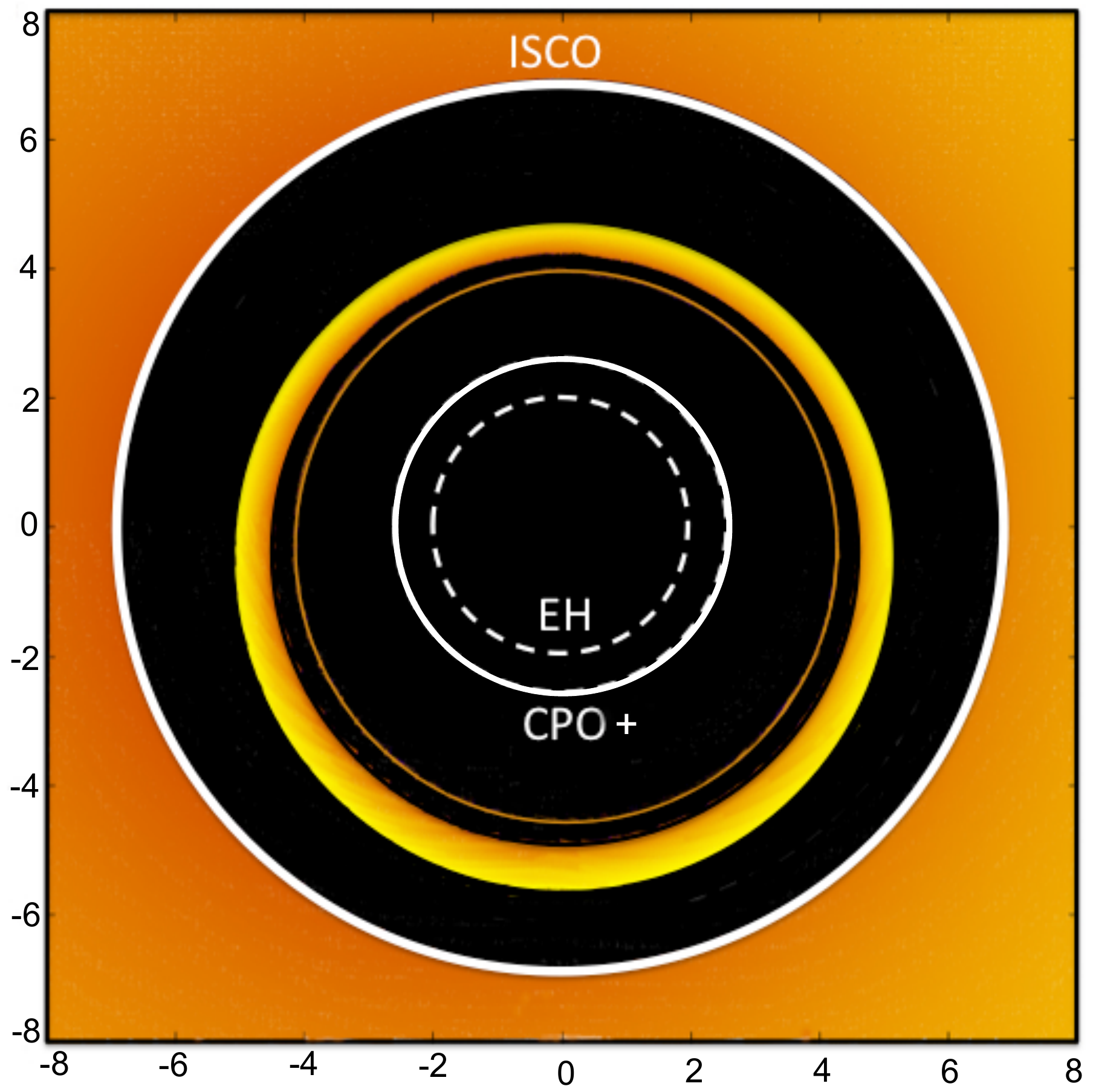, scale=0.3}
\hspace{+0.2cm}
\psfig{figure=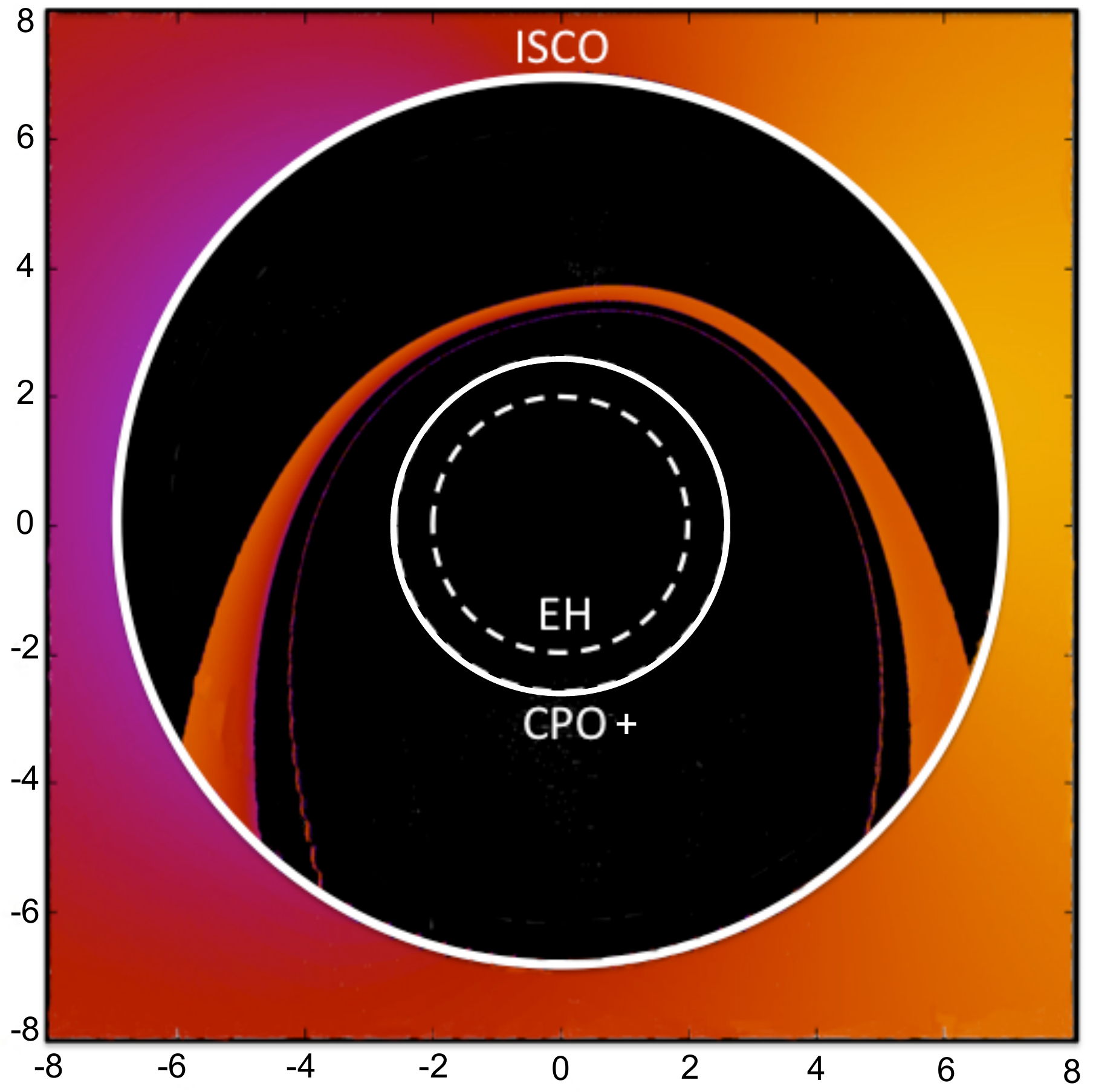, scale=0.3}
\hspace{+0.2cm}
\psfig{figure=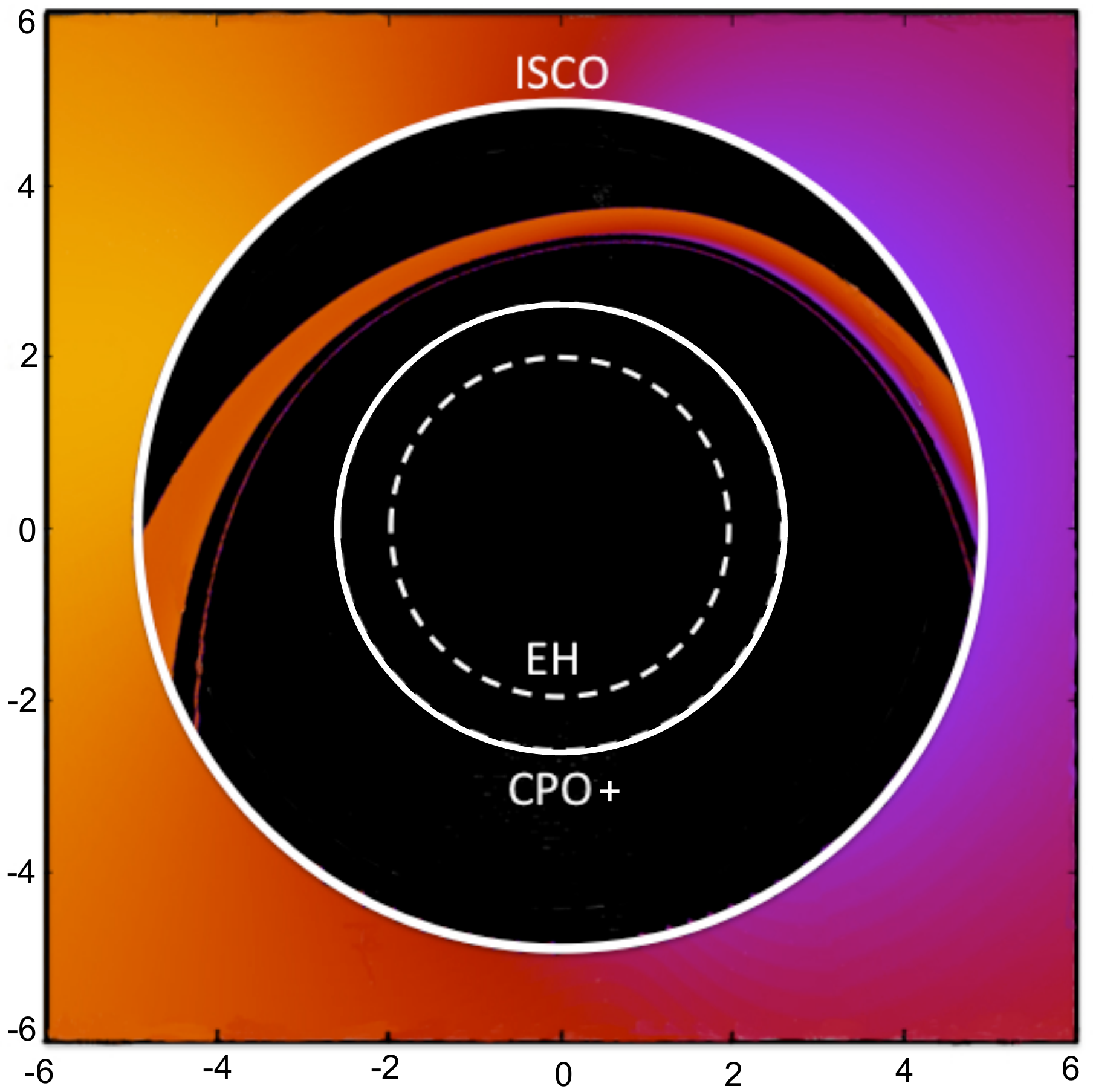, scale=0.3}
}
\caption{Examples of the plunging region (black area) between the inner disk and the BH at the emission plane map. The structures inside the plunging regions represent the locus of the points where the HOI rays cross the equatorial plane. The left and central panels refer to a counter-rotating disk with spin $a=-0.3$ and inclination angles of $i=10^{\circ}$ and $i=60^{\circ}$, respectively. The right panel illustrates the case of a corotating disk with spin $a=0.3$ and inclination angle of $i=60^{\circ}$. Colors represent the frequency shift $g$ of the rays. The outer continuous circle represents the ISCO, whereas the inner continuous circle shows the CPO+ radius. The innermost dashed circle is the event horizon.}
\label{fig:figHOI}
\end{figure*}

Figure \ref{fig:figHOI} shows the emission plane maps displaying the points where the HOI rays cross the plunging region. It is apparent that their location depends strongly on the observer inclination angle.

\subsection{Emissivity profiles in the lamp-post geometry}
\label{sec:lampsimulations} 
An often adopted prescription for the disk radial emissivity profile is that of 
the so-called lamp-post model, which consists of a point source illuminating the disk from above (and below) and located on the BH spin axis. This geometry is meant to approximate the illumination produced by a compact inner disk corona or a jet. The lamp-post height is the crucial parameter which determines the radial dependence of the disk emissivity, whereas the lamp-post spectrum and luminosity drive the ionisation state of matter and contribute determining the local reflected spectrum, including the Fe line \citep[e.g.][]{Miniutti2003, Wilkins2012}.

\begin{figure}
\centering
\hbox{
\psfig{figure=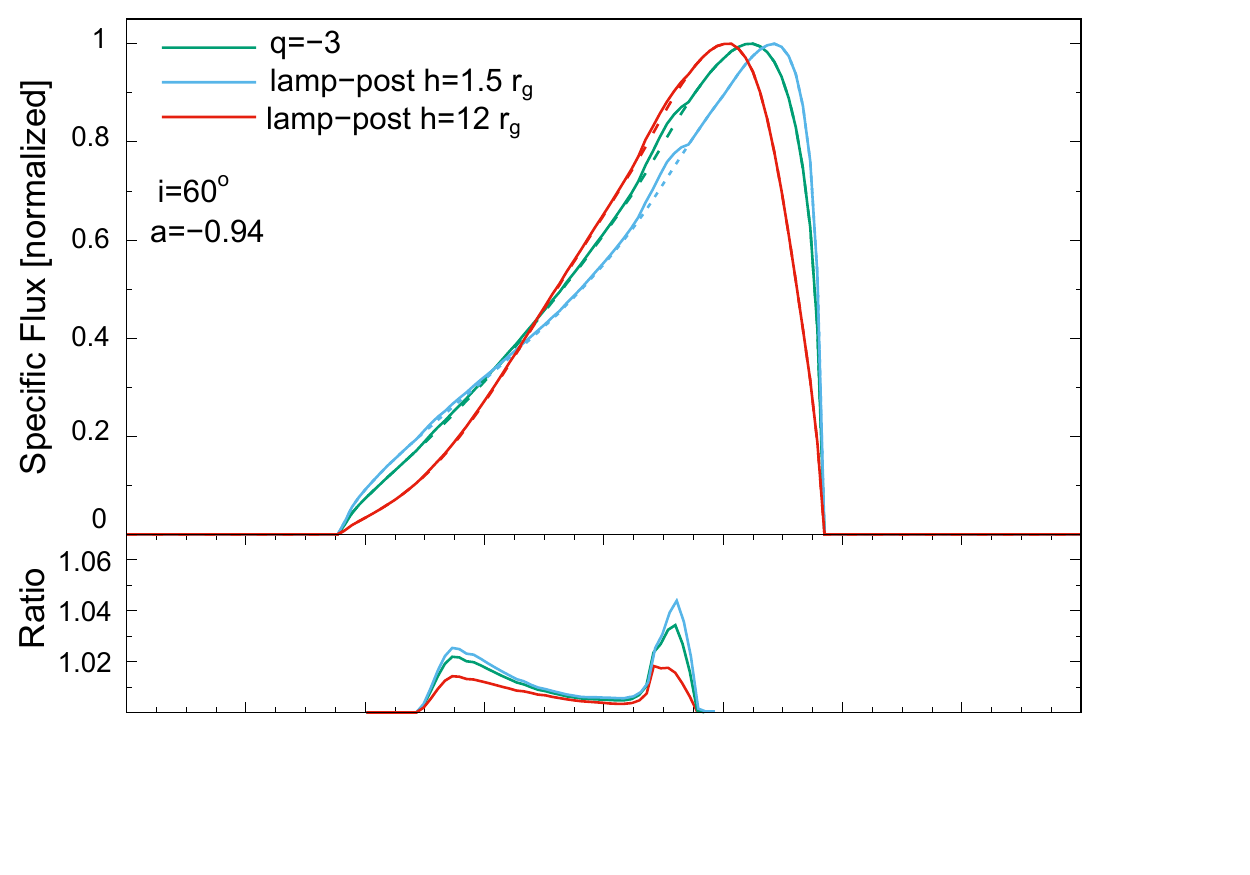, scale=0.73}
}
\vspace{-0.1cm}
\hbox{
\psfig{figure=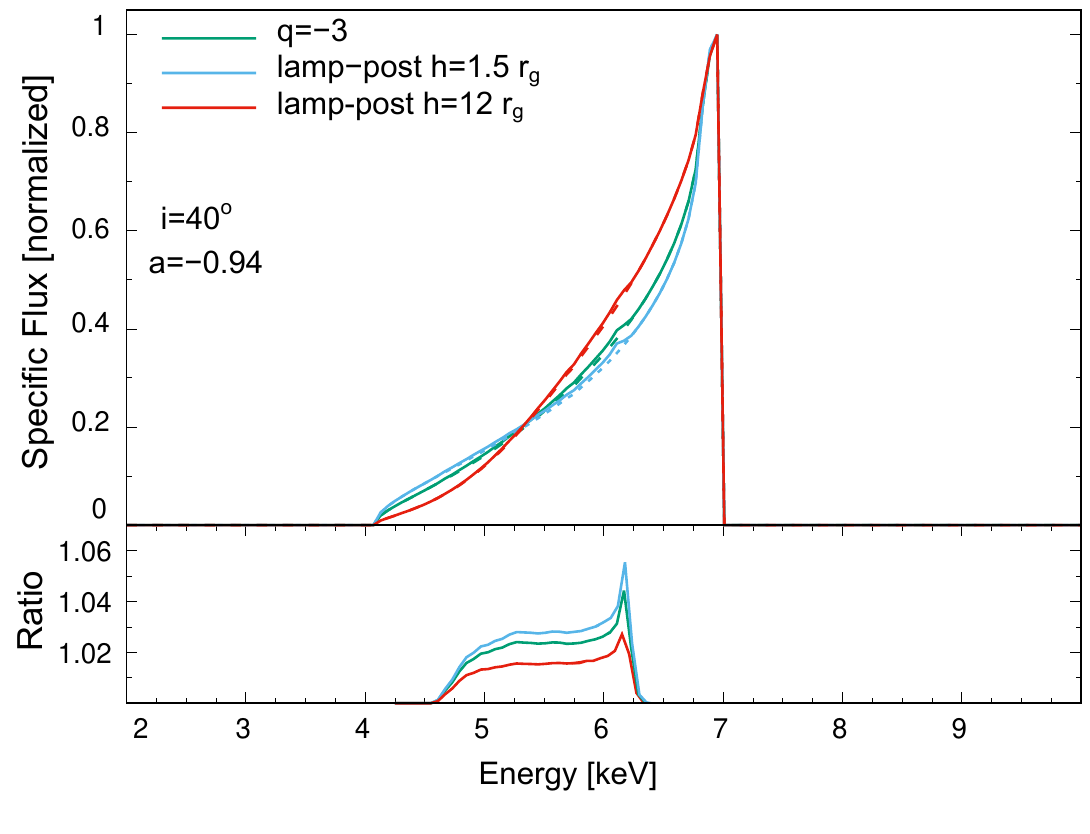, scale=0.73}
}
\caption{Comparison of Fe $K_\alpha$ line profiles calculated for different disk emissivity models with (solid lines) and without (dashed lines) disk HOIs contributions along with their ratios (bottom panels). The cyan and red profiles are for the lamp-post emissivity model of \citet{Kammoun2019} with lamp-post height $h=1.5\,r_g$ and $h=12\,r_g$. The green profiles are for a power law  emissivity profile $\mathcal{R}(r) \propto r^{-3}$ (green). The disk is counter-rotating with spin parameter fixed at $a=-0.94$; the observer inclination angle is $i=60^{\circ}$ (top panel) and $i=40^{\circ}$ (bottom panel).}
\label{fig:lamp-post}
\end{figure}

To work out some examples we adopted the effective emissivity profile calculated by \citet[see Fig. 2 in][]{Kammoun2019}\footnote{For this we used two suitably defined twice-smoothly-broken power laws as radial emissivities in Eq. (\ref{ebin_flux}).} for lamp-post heights of $12~\mathrm{r_g}$ and $1.5~\mathrm{r_g}$ and BH spin of $a=0.94$.
In the $12~\mathrm{r_g}$ case the emissivity has a flatter trend than a $q=-3$ power law up to radii comparable to the lamp-post height, as expected from simple geometrical arguments.
Instead, when the lamp-post is very close to the BH at $1.5~\mathrm{r_g}$, the emissivity profile is steeper than a $q=-3$ power law for radii $\lesssim 10~\mathrm{r_g}$, as a result of very large photon deflection angles.
Figure \ref{fig:lamp-post} shows representative Fe-line profiles along with HOIs contributions that we calculated for both lamp-post heights and selected inclination angles in the case of a counter-rotating disk (in the co-rotating case the HOI contribution is negligible due to the small ISCO radius). A small lamp-post height $h=1.5~\mathrm{r_g}$ slightly boosts ($ \lesssim 1\%$) the HOIs contribution with respect to the power-law emissivity profile $\mathcal{R}(r) \propto r^{-3}$, since the inner part of the disk, which gives rise to the strongest of the disk HOIs, is more illuminated. On the contrary, the large lamp-post height $h=12~\mathrm{r_g}$ suppresses ($ \lesssim 2\%$) the HOI contribution.

\section{spectral simulations of the black hole X-ray binary GRO J1655-40} 
\label{sec:data}    

In this section we investigate the way in which the HOI contribution to the iron line profile affects the determination of the system parameters such as BH spin and inner disk radius, when the line profile is fitted with a model that does not include HOIs.
Since the signal-to-noise ratio of current instrumentation is insufficient for this purpose, we chose to simulate observations with next generation large area instrumentation such as that planned for {\it eXTP} \citep[][]{Zhang2019}. Its Large Area Detector (LAD) will afford an effective area of $\sim3$~m$^2$ in the $\sim$8~--~10~keV range, about $5-10$ times larger than that of the largest X-ray instruments of the past and present generations, as well as {\it Athena}'s \citep{Athena}. Using the {\it eXTP} public response matrices (version 4, released in 2017) we simulated LAD spectra in the nominal energy range $2-30$ keV with \textsc{Xspec} version 12.10.1s \citep{Arnaud1996}. 
In the spectral analysis, we also considered a systematic uncertainty on the LAD background knowledge of 0.5\%.

We adopted as a basis the observed parameters of GRO J1655-40, a BH low mass X-ray binary among the closest to Earth. Its distance, initially thought to be about 3.2 kpc, was later estimated to be much lower, with an upper limit at 1.7 kpc \citep{Foellmi2009}. The many works cited by Table 1 in \cite{Foellmi2009} also provide estimates of the BH mass in the $\sim$4 - 8 M$_{\odot}$ range: therefore, we will assume a BH mass of $\mathrm{6~M_{\odot}}$, corresponding to an Eddington luminosity of $L_{\mathrm{Edd}} \sim 7.56 \times 10^{38}~\mathrm{erg~s^{-1}}$, and convert fluxes to luminosities assuming a distance of 1.7 kpc. Its X-ray spectra have displayed strong, broad Fe line profiles both in the soft and hard source states \citep[see e.g.,][]{BalucinskaChurch2000, DiazTrigo2007, Takahashi2008, Reis2009}. 

Since our interest lies mainly in gauging the effects of the HOI contribution to the \fekal, we adopted a simple model composed of \texttt{TBabs}*(\texttt{pexriv} + \texttt{diskbb} + \texttt{relline} + \texttt{HOIeffects}) to approximate the continuum spectral components (plus the Fe line) from \gro\ in its rising low hard state and similarly behaving BH X-ray binaries. \tbabs\ accounts for the X-ray absorption by the interstellar medium \citep{Wilms2000}, \pexriv\ describes the exponentially cutoff power law and the reflection component coming from the disk except for emission lines \citep{Magdziarz1995}, \diskbb\ adds the disk multi-blackbody emission \citep[e.g.,][]{Mitsuda1984} and \relline\ generates an Fe $K_{\alpha}$ line profile in Kerr spacetime \citep{Dauser2010}. Finally by \texttt{HOIeffects} we indicate the added contribution of the HOIs to the Fe $K_{\alpha}$ line profile calculated with the same disk parameters used by \relline.
Through \textsc{Xspec}'s \texttt{fakeit} command we simulated the observation of such a system for various disk configurations, then examining the impact of the HOIs on the spectrum and on the determination of the system parameters.

A caveat is in order here. In principle, all the reflection components originating from the disk are affected by the same relativistic effects that give rise to the broad, asymmetric Fe-line profile, since they are expected to be produced in the same area: for this reason many current models convolve the whole reflected spectrum from the disk with a single relativistic broadening shape. A bias in the estimate of the line parameters may be introduced by not accounting for the broadening of the continuum \citep[see e.g.][]{Reynolds2020}. However, a different issue has been recently found, namely that convolution routines within \textsc{Xspec} give rise to spurious effects when steep features of the kernel function encounter narrow spectral components \citep[][]{LaPlaca2021}. Since in our case the amplitude of such effects turned out to be larger than the contribution of the HOIs to the line profile discussed in Sect. \ref{sec:simulations}, we elected to adopt a model consisting of a single, relativistically broadened Fe K$\alpha$ line superposed on a non-relativistic reflection continuum.

\begin{table}
\centering
\caption{Starting parameters that are the same across all the simulated observations. The additional fitting parameters are the inner disk radius $r_{in}$, the BH spin $a$, the inclination $i$, the normalization of the \pexriv\ and \relline\ components, and the ones linked to them. Parameters indicated by the letter $f$ are kept fixed during the subsequent analysis, and the radial emissivity is kept uniform over the whole disk.
$^{\dagger}$The angular emissivity is considered isotropic.}
\begin{tabular}[c]{lLCC}
		\hline
		 Models   & $ Parameters $ & $Values $ & $Typical best-fit value$  \\
		\hline 
		\tbabs      & {\rm N_H}\ (10^{22}\ \mathrm{cm}^{-2})     &  0.7   &  0.700 \pm 0.004\\
		\pexriv     & \Gamma  &  1.8          &   1.801 \pm 0.003 \\   
		      & \mathrm{foldE\ (keV)}     &  100          &  f \\          
		      & {\rm rel_{refl}}&  0.5           &   0.503 \pm 0.006 \\    
		      & \mathrm{Redshift} &  0.0           &   f\\   
		      & \mathrm{abund}    &  1.0           &  f \\ 
		      & \mathrm{Fe~abund}  &  1.0           &  f \\  
		      & \mathrm{T_{disk}}\ (10^6\ \mathrm{K})  &  1 &  f \\          
		      & \xi\ (\mathrm{erg\ cm\ s^{-1}})       &  500 &   501 \pm 5 \\          
		\diskbb     & \mathrm{T_{in}}\ (\mathrm{keV})       &  0.5 &  0.4999 \pm 0.0005 \\ 
		\relline    & \mathrm{LineE}\ (\mathrm{keV})     &  6.4 &   6.40 \pm 0.02 \\         
		     & \mathrm{Index1}    &  3.0          &  2.98 \pm 0.02 \\         
		     & \mathrm{R_{br}}\ (\mathrm{r_{\rm g}})       &  100 &  f  \\
		     & \mathrm{R_{out}}\ (\mathrm{r_{\rm g}})      &  900 &  f \\  
		     &    z         &  0.0 &  f  \\         
		     & \mathrm{limb}&  0.0^{\dagger} &  f \\
		\hline 
	\end{tabular}
    \label{tab:modelpar}
\end{table}

The model parameters that were initially set to the same values across all simulations are listed in Table \ref{tab:modelpar}. The values for these parameters were chosen by comparing different archival analyses of this source \citep[such as, e.g.,][]{Shaposhnikov2007, CaballeroGarcia2007, Takahashi2008, Kalemci2016}, and are representative of similar BH X-ray binaries. We set the normalizations of \pexriv\ and \relline\ so that the 2 -- 10~keV flux would always be $\sim1.3\times10^{-9}~\mathrm{erg~cm^{-2}~s^{-1}}$, corresponding to a broad-band X-ray luminosity of $\sim 0.002~L/L_{\rm Edd}$, while keeping the equivalent width of the Fe $K_{\alpha}$ line profile close to 0.15 keV in all cases. 

We used the \texttt{fakeit} command within \textsc{Xspec} to simulate 500~ks exposure spectra from twenty-one different configurations, five with inclination $i = 20^{\circ}$, eight with  $i = 40^{\circ}$, and eight with $i = 60$\textdegree; we simulate each case five times, for a total of 105 simulations.
Note that a fitting model component which includes the HOIs' profile has not yet been developed.
Therefore we adopted in the fit the same model used to generate the simulated spectra, having removed the HOI component: this way we can determine the biases introduced by not accounting for the HOIs of the \fekal\footnote{Before fitting we decoupled \diskbb's normalization and \relline's $r_{\rm in}$ and set the latter to be computed in units of $r_{\rm ISCO\pm}$: this was done to avoid possible inconsistencies between the value of the inner disc radius as determined by \textsc{Xspec} and the one used internally by \relline\ when $r_{\rm in}$ is given in units of $r_g$.}.

\begin{figure*}
\centering
\hbox{
\psfig{figure=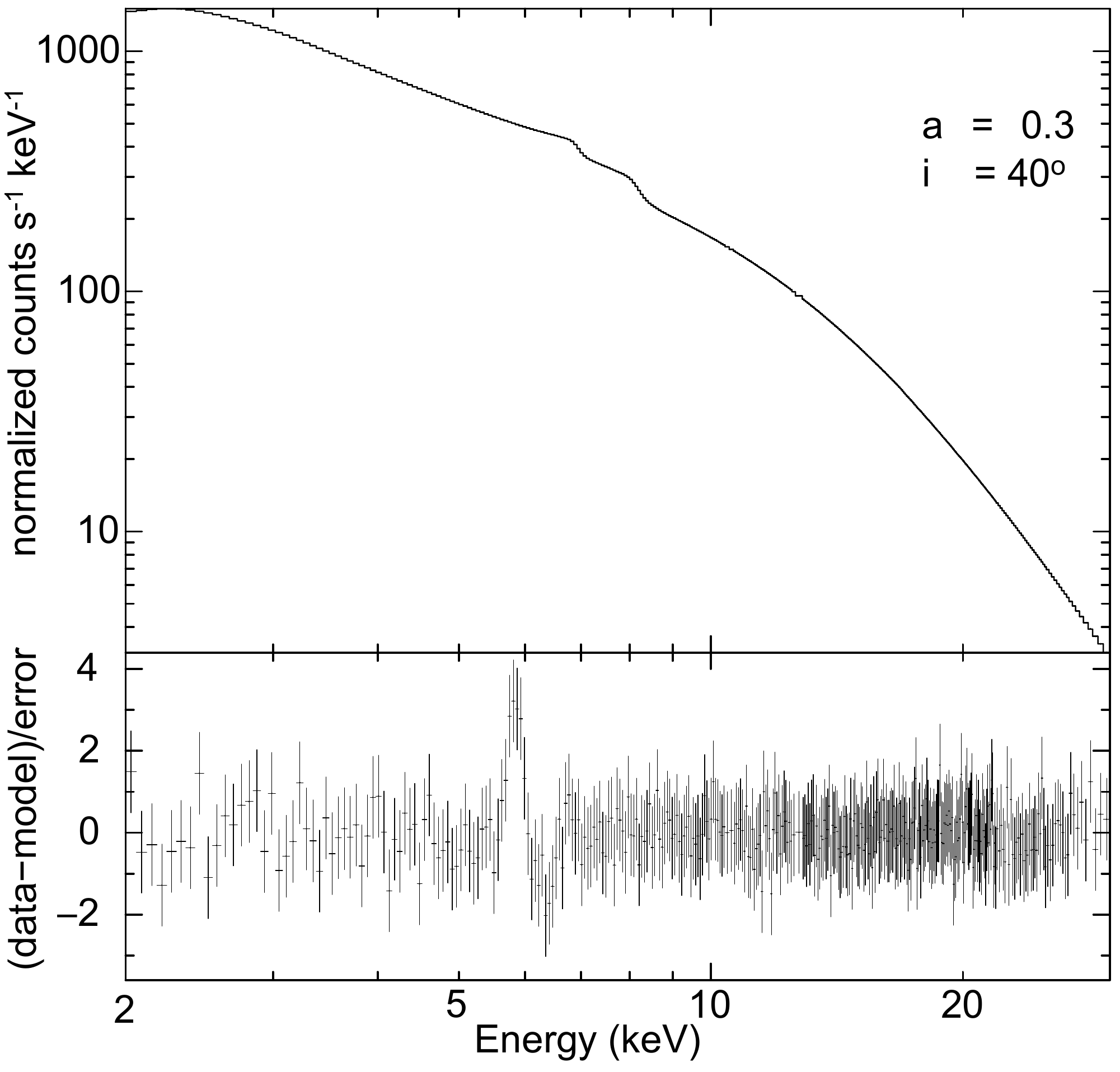, scale=0.42}
 \hspace{0.15cm}
\psfig{figure=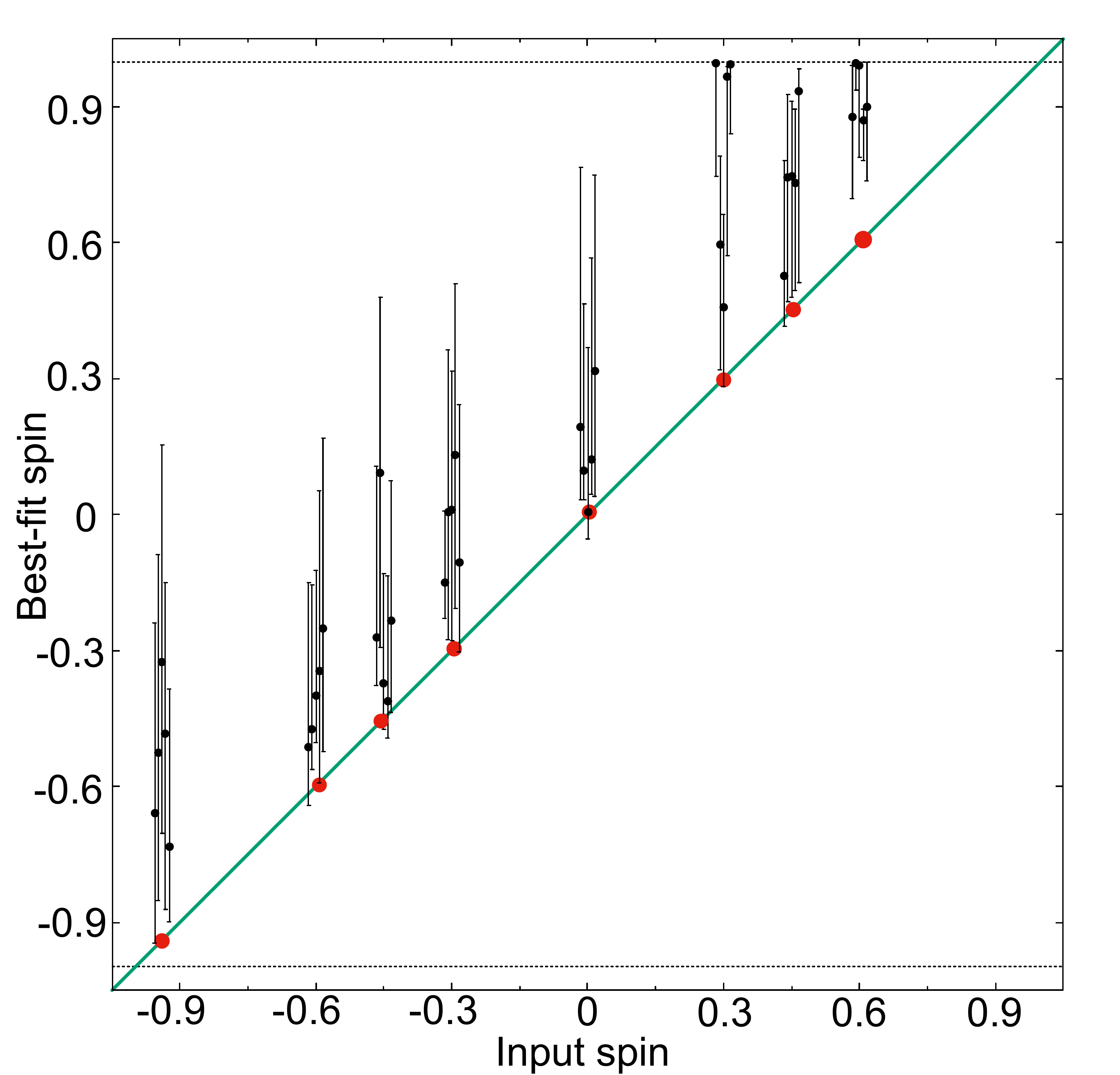, scale=0.25}
}
\caption{ \textit{Left side - }Simulated eXTP/LAD spectrum of \gro\ in a low state with \texttt{TBabs}*(\texttt{pexriv} + \texttt{diskbb} + \texttt{relline} + \texttt{HOIeffects}) best-fit model. The lower panel shows the residuals between the data and model: this  $i = 40$\textdegree, $ a = 0.3$ case is among the few in which the effects of the HOIs are still visible after the error calculation.  \textit{Right side - }Distribution of the best-fit values of the spin for the 40 simulations with $i = 60$\textdegree, with their respective $95\%$-confidence intervals. Exclusively for plotting purposes, the points are slightly spread horizontally to avoid the overlap of their confidence intervals: the quintuplets of points are to be considered to lie, left to right, at $a_{\mathrm{input}} = -0.94, -0.6, -0.45, -0.3, 0.0, 0.3, 0.45, 0.6$. The green diagonal represents the $a_{\mathrm{best-fit}} = a_{\mathrm{input}}$ line which all points' confidence intervals should cross, were they compatible with the injected values: it is quite evident how the spin is consistently overestimated during the analysis. The two horizontal dashed lines represent \relline's limits on the spin parameter at $a=-0.998$ and $a=0.998$.}
\label{fig:fig7}
\end{figure*}

During our spectral fits, notwithstanding the changes in the remaining parameters, the values of almost all parameters of the continuum and their confidence intervals were stable, the former being always consistent with the starting values to within the respective uncertainties (here ``continuum'' indicates the combination of \tbabs, \pexriv\ and \diskbb). In all cases reduced $\chi^2$ values less than 1 were obtained in the fits, as a result of the systematic uncertainty on the background we introduced. The last column in Table \ref{tab:modelpar} reports typical best-fit values for the parameters that are shared among all simulations. The best-fit values and their associated $95\%$-confidence intervals on the two main parameters of interest, the spin $a$ and inner disk radius $r_{\rm in}$, are given in Table \ref{tab:simul}, where we show the best-fit values for one simulation out of the five for each case, randomly selected. As expected the HOIs' signatures are in most cases made unrecognisable in the residuals by the leveling action resulting from changes in the other parameters (see Fig. \ref{fig:fig7}, left subfigure, for one among the handful of examples that instead do show some residuals). However, this often brings some of the line parameters, especially the spin, to values which are incompatible with the initial ones even with rather large $95\%$-confidence intervals: the other parameters that are sometimes rendered incompatible are the line energy, inclination, and emissivity index, although only by a few percents. 
\begin{table}
    \centering
\caption{Best-fit value of the two main parameters of interest, $ r_{in} $ and $ a $, with their $95\%$-confidence intervals.}
    \renewcommand{\arraystretch}{1.5}
    \begin{tabular}[c]{CCCCC}
    \hline
    \multicolumn{2}{C}{$Input parameters$} & \multicolumn{3}{C}{$Best-fit parameters$}  \\
    \hline
    i & a  & r_{in}~(r_{\rm ISCO\pm}) & a & \mathrm{\#~biased~sims} \\
    \hline
    20$\textdegree$ &  0.6  & 1.13_{-0.13}^{+0.23}    &  0.74_{-0.13}^{+0.12} & 1 \\
    20$\textdegree$ &  0.3  & 1.27_{-0.14}^{+0.09}    &  0.65_{-0.14}^{+0.07} & 4 \\
    20$\textdegree$ &  0.0  & 1.09_{-0.09}^{+0.15}    &  0.19_{-0.20}^{+0.20} & 3 \\
    20$\textdegree$ & -0.3  & 1.48_{-0.16}^{+0.26}    &  0.48_{-0.36}^{+0.35} & 1 \\
    20$\textdegree$ & -0.6  & 1.00_{-0.00}^{+0.27}    & -0.62_{-0.05}^{+0.72} & 1 \\
    40$\textdegree$ &  0.6  & 1.00_{-0.00}^{+0.05}    &  0.65_{-0.02}^{+0.05} & 3 \\
    40$\textdegree$ &  0.45 & 1.00_{-0.00}^{+0.02}    &  0.49_{-0.03}^{+0.06} & 3 \\
    40$\textdegree$ &  0.3  & 1.00_{-0.00}^{+0.03}    &  0.37_{-0.02}^{+0.06} & 4 \\
    40$\textdegree$ &  0.0  & 1.00_{-0.00}^{+0.27}    &  0.06_{-0.03}^{+0.19} & 3 \\
    40$\textdegree$ & -0.3  & 1.00_{-0.00}^{+0.11}    & -0.28_{-0.03}^{+0.28} & 0 \\
    40$\textdegree$ & -0.45 & 1.00_{-0.00}^{+0.14}    & -0.47_{-0.09}^{+0.34} & 0 \\
    40$\textdegree$ & -0.6  & 1.00_{-0.00}^{+0.17}    & -0.68_{-0.12}^{+0.47} & 0 \\
    40$\textdegree$ & -0.94 & 1.05_{-0.05}^{+0.71}    & -1.00_{-0.00}^{+1.25} & 0 \\
    60$\textdegree$ &  0.6  & 1.44_{-0.35}^{+0.30}    &  0.87_{-0.09}^{+0.03} & 5 \\
    60$\textdegree$ &  0.45 & 1.29_{-0.29}^{+2.15}    &  0.73_{-0.24}^{+0.16} & 4 \\
    60$\textdegree$ &  0.3  & 2.66_{-0.70}^{+1.12}    &  0.97_{-0.40}^{+0.02} & 4 \\
    60$\textdegree$ &  0.0  & 1.01_{-0.01}^{+0.22}    &  0.12_{-0.08}^{+0.44} & 4 \\
    60$\textdegree$ & -0.3  & 1.20_{-0.20}^{+0.46}    &  0.13_{-0.34}^{+0.38} & 4 \\
    60$\textdegree$ & -0.45 & 1.00_{-0.00}^{+0.14}    & -0.41_{-0.08}^{+0.28} & 3 \\
    60$\textdegree$ & -0.6  & 1.09_{-0.09}^{+0.28}    & -0.34_{-0.25}^{+0.40} & 4 \\
    60$\textdegree$ & -0.94 & 1.19_{-0.17}^{+0.20}    & -0.48_{-0.39}^{+0.33} & 5 \\
    \hline
    \end{tabular}
    \label{tab:simul}
\end{table}

Of the 105 simulations we ran, 85 yielded best-fit spins that overestimate the input values. This is the most evident effect of not including the HOIs in the fitting model; it is most pronounced in systems observed under 60\textdegree\ inclination and/or co-rotating disks. 
The last column in Table \ref{tab:simul} indicates how many simulations out of five for each case presented values of the spin parameter whose $95\%$-confidence intervals were incompatible with the input ones: they amount to a total of 56 simulations, with spin values always higher than the input ones. The most extreme instance is the $i = 60$\textdegree, $ a = 0.6 $ case, four simulations of which have their upper limit for the spin parameter only constrained by the model's highest possible value, $ a = 0.998$. Fig. \ref{fig:fig7}, right panel, shows the distribution of the best-fit values of the spin in the $i=60$\textdegree\ simulations versus their input value, providing the most striking example of the consistent overestimation of the spin. 
In general, low-inclination and negative-spin simulations show the largest uncertainties on the best-fit spin, with confidence intervals that can almost span the entire range for strongly counter-rotating disks observed under $i=20$\textdegree. Conversely, the co-rotating cases with $i=40$\textdegree\ return the narrowest confidence intervals, possibly because of the very sharp shape of the HOIs. 

The second most-common effect of not accounting for the HOIs in the \fekal\ is a small decrease, in the order of a few percent, of the radial emissivity index: this makes the parameter incompatible with its input value in more than thirty simulations. Other secondary effects include minor variations of the inclination and rest line energy with respect to their input values, mostly increases for the former parameter and decreases for the latter. Although all these deviations are almost always occurring during analyses in which the spin is wrongly determined too, we found a handful of exceptions: most notably, in the $i = 40$\textdegree, $ a = -0.3 $ case all five simulations showed underestimates of both the line energy and emissivity index, together with constant overestimates of the inclination, in spite of the fact that the spin was always compatible with the actual one to within its confidence intervals.

Finally, as Table \ref{tab:simul} shows, in most of the cases featuring a spin overestimate the best-fit inner disk radius is larger than the expected ISCO: this is understandable, since the radius of the ISCO itself shrinks with rising spin. This also confirms that the high spin values we found are driven by the overall profile of the line, rather than merely by the reddest part of the red wing, as determined by the ISCO radius.

These results demonstrate that ignoring HOI effect may introduce a bias towards higher values of the BH spin $a$ in the analyses of systems with a wide range of parameters, provided that the plunging region is optically thin and that the signal-to-noise ratio of the data is high enough.

To assess, albeit approximately, the significance of the non-modelled HOI effects, we compared the best-fit model in our analyses above with a fit carried out with the full model used while generating the simulated spectra, i.e., \texttt{TBabs}*(\texttt{pexriv} + \texttt{diskbb} + \texttt{relline} + \texttt{HOIeffects}): as stated earlier, a HOI-fitting model component has not yet been developed, so \texttt{HOIeffects} is the same fixed HOI shape inputted in generating each simulation. This recovered parameter values in agreement with the input ones and also yielded better fits to the simulated spectra ($\chi^2$ decreases in the $\sim3-50$ range were found, depending on the HOI strength and shape), even though the improvement is not apparent from the residuals except for only a few cases: the simulation corresponding Fig. \ref{fig:fig7}, left panel, represents the most extreme example with $\Delta\chi^2$=55.

\section{DISCUSSION} 
\label{sec:summary} 
We  analysed the contribution of HOI rays to the relativistic Fe-line profile emitted from an accretion disk around a Kerr BH, over a range of BH spin parameters, disk inclinations and by considering different disk radial emissivity profiles. Concerning the radial dependence of the optical depth of the plunging region in the Kerr metric, in addition to the standard estimate we considered a lower limit on it for the case in which disk matter is still torqued well inside the ISCO radius. Like \citet{BeckDone2004} and \citet{Bambi2020} we assumed in our calculations that the whole plunging region is optically thin so that HOI rays can travel unimpeded through it. Characteristics HOI features are imprinted in the profile extending from the red wing up to energies somewhat lower than those of the blue peak of the line. They typically contribute $\sim 1 - 4$\% to the specific flux of the profile, with peaks in the $\sim 5 - 7$\% range in  several intermediate-inclination cases. 
Our analysis concentrated on highly negative ($a=-0.995$) to moderately large positive values ($a=0.6$) of the BH spin parameter $a$ and evidenced that the relative contribution from the HOIs is higher for intermediate-low inclinations ($20^{\circ}$ and $40^{\circ}$) and decreases for
increasing values of $a$, as a result of the ISCO approaching the CPO+, thus reducing the size of the plunging region. The profiles and trends that we obtained are similar to those discussed by \citet{Bambi2020} over the common ranges of $a$ and $i$\footnote{Judging by the profiles they display, we suspect that their outer disc radius was set to a lower value than ours; this slightly changes the relative strength of the HOIs}.  

The overall HOI contribution to the Fe-line flux, an effect in the $\sim 0.3 - 1.4$\% range, would be virtually impossible to reveal with present X-ray instrumentation. We simulated the effect of the HOIs in the X-ray spectrum of the BH X-ray binary system \gro\ in a low hard rising state as studied with the instrument LAD onboard of the planned X-ray astronomy mission eXTP in $\sim 500$~ks observations. We then fitted the simulated spectra by using a Fe-line model that did not include the emission feature from HOIs. The input values of some parameters, such as the line rest energy and disk inclination, were not correctly recovered in some of these fits. Significant differences were found in the determination of the spin parameter $a$, with consistently higher values in the majority of our simulations. By inference we conclude that not accounting for the contribution of the HOIs to the line profile would lead to a bias in BH spin estimates.

\citet{Bambi2020} report that they did not find any appreciable deviation in their simulations between spectra in which HOIs were or were not included.
We note that they simulated the spectra of an AGN a factor of about 10 fainter than our simulations of \gro, and observed  with the X-ray Integral Field Unit (X-IFU) instrument planned for Athena \citep{Athena}, whose effective area at $\sim 6$~keV is $<1/10$ that of eXTP.
Moreover they considered only one value of inclination ($i=70^{\circ}$) and large values of BH spin ($a = 0.70$ and 0.95), for which the effects of the HOIs is less pronounced owing to the shrinking ISCO radius. The model they use during the analysis also includes the deformation parameter $a_{13}$ of the Johannsen metric to study possible deviations from general relativity: it is not clear whether this parameter may have introduced additional degeneracies \citep[but see also Fig. 4 and 5 in][]{Bambi2020}. We suspect that a combination of the above factors allowed the influence of the HOIs to go undetected in the \citet{Bambi2020} simulations. Further effects that could have contributed to this are currently under study.

We conclude that with next-generation, very-large-area X-ray instrumentation the effect of an extreme gravitational lensing phenomenon such as the HOI contribution to the Fe-line profile should be appreciable in long observations of BHs in galactic X-ray binaries, provided they accrete at sufficiently low rates ($\lesssim 0.01\, L_{\rm Edd}$) that most of the plunging region remains optically thin. As we saw in Sect. \ref{sec:data}, the HOI spectral signature would often be masked during the analysis by variations of the other parameters while still providing good fits to the data.
In any case the development of suitable fitting models which include the effects of the HOIs will be key in removing biases and searching for such features in very-high signal to noise spectra of the future.

The characteristics of relativistic Fe-line profiles from accretion disks have been studied also in some alternative gravity theories and proposed as a means of testing their predictions against those of GR \citep[see e.g.][and references therein]{Johannsen2013,Bambi2016}. Our results demonstrate that taking into account the HOI contributions in the pure BH-Kerr-metric approach discussed here would be essential also in this context, to avoid mistaking their effects for departures from the predictions of GR.

Finally we note that HOIs in their own right and likewise other extreme lensing phenomena \citep[see, e.g.,][]{Hioki2009,Bambi2011,Broderick2014} are amenable to generalisations to other theories, including {\it e.g.}: modified Kerr-like spacetimes with arbitrary multipole moments, which may falsify the Kerr hypothesis and thus GR itself \citep{Yagi2016,Konoplya2016}; violations of the Kerr bound relation ($a \leq 1$) that may prove the presence of a naked singularity \citep{Penrose1969,Shapiro1991}; Kerr-like brane-world or super-spinning spacetime geometry \citep{Aliev2005,Horava2009}. Future works may succeed in singling out Fe-line HOI diagnostics also in the context of such theories. 

\section*{Acknowledgements}
The authors would like to thank the referee, T. Dauser, who provided useful and detailed comments to improve the manuscript. PB is grateful to the International Space Science institute (ISSI) in Bern for their local support and hospitality, where part of this work has being carried out. 
VDF thanks Gruppo Nazionale di Fisica Matematica of Istituto Nazionale di Alta Matematica (INDAM) for support.  LS and ADR acknowledge financial contribution from ASI-INAF agreements 2017-14-H.O. LS acknowledges financial contribution from ASI-INAF agreement I/037/12/0 and from ``iPeska'' research grant (P.I. Andrea Possenti) funded under the INAF call PRIN-SKA/CTA (resolution 70/2016). LS and RLP acknowledge financial contribution from ``PRIN INAF 2019 n.15'' (P.I. T. Belloni). 

\section*{Data Availability}

The data underlying this article will be shared on reasonable request to the corresponding author.

\bibliographystyle{mnras}
\bibliography{references}
\label{lastpage}

\end{document}